\documentclass[manuscript]{aastex}

\newcommand{\sodium}{Na~{\sc i}}
\newcommand{\potassium}{K~{\sc i}}
\newcommand{\oi}{O~{\sc i}}
\newcommand{\caii}{Ca~{\sc ii}}

\shorttitle{Na and K Lines}
\shortauthors{Pascucci et al.}

\begin{document}


\title{Narrow Na and K Absorption Lines Toward T Tauri Stars –– Tracing the Atomic Envelope of Molecular Clouds}

%\title{1. A Cloud Origin for the Sharp Absorption in the Na and K Resonance Lines of T Tauri Stars OR 2. The different distribution of atomic and molecular gas in the Taurus star-forming %region OR….}


\author{I. Pascucci}
\affil{Lunar and Planetary Laboratory, The University of Arizona, Tucson, AZ 85721, USA}
\email{pascucci@lpl.arizona.edu}

%\and

\author{S. Edwards}
\affil{Five College Astronomy Department, Smith College, Northampton, MA 01063, USA}

%\and

\author{M. Heyer}
\affil{Department of Astronomy, University of Massachusetts, Amherst, MA 01003-9305, USA}

%\and

\author{E. Rigliaco}
\affil{Institute for Astronomy, ETH Zurich, Wolfgang-Pauli-Strasse 27, CH-8093 Zurich, Switzerland}

%\and

\author{L. Hillenbrand}
\affil{Department of Astronomy, California Institute of Technology, Pasadena, CA 91125, USA}

%\and

\author{U. Gorti\altaffilmark{1}}
\affil{SETI Institute, Mountain View, CA 94043, USA}
\altaffiltext{1}{NASA Ames Research Center, Moffett Field, CA 94035, USA}

%\and

\author{D. Hollenbach}
\affil{SETI Institute, Mountain View, CA 94043, USA}

\and

\author{M. N. Simon}
\affil{Lunar and Planetary Laboratory, The University of Arizona, Tucson, AZ 85721, USA}

%\author{C. D. Biemesderfer\altaffilmark{4,5}}
%\affil{National Optical Astronomy Observatories, Tucson, AZ 85719}
%\email{aastex-help@aas.org}

%\and

%\author{R. J. Hanisch\altaffilmark{5}}
%\affil{Space Telescope Science Institute, Baltimore, MD 21218}


%\altaffiltext{1}{Visiting Astronomer, Cerro Tololo Inter-American Observatory.
%CTIO is operated by AURA, Inc.\ under contract to the National Science
%\altaffiltext{2}{Society of Fellows, Harvard University.}
%\altaffiltext{3}{present address: Center for Astrophysics,
%    60 Garden Street, Cambridge, MA 02138}
%\altaffiltext{4}{Visiting Programmer, Space Telescope Science Institute}
%\altaffiltext{5}{Patron, Alonso's Bar and Grill}


% MH - the abstract is rather long and could be condensed by removing some of the details

\begin{abstract}

We present a detailed analysis of narrow \sodium{} and \potassium{} absorption resonance lines toward nearly 40 T Tauri stars in Taurus with the goal of clarifying their origin.
The \sodium{} 5889.95\,\AA{} line is detected toward all but one source, while the weaker \potassium{} 7698.96\,\AA{} line in about two thirds of the sample. The similarity in their peak centroids and the significant positive correlation between their equivalent widths demonstrate that these transitions trace the same atomic gas. 
The absorption lines are present towards both disk and diskless young stellar objects, which 
 excludes cold gas within the circumstellar disk as the absorbing material. 
A comparison of \sodium{} and CO detections and peak centroids demonstrates that the atomic and molecular gas are not co-located, the atomic gas is more extended than the molecular gas. The width of the atomic lines corroborates this finding and points to atomic gas about an order of magnitude warmer than the molecular gas.
The distribution of \sodium{} radial velocities shows a clear spatial gradient along
the length of the Taurus molecular cloud filaments. This suggests that absorption is associated with the Taurus molecular cloud. 
Assuming the gradient is due to cloud rotation, the rotation of the atomic gas
is consistent with differential galactic rotation while the rotation of the molecular
gas, although with the same rotation axis, is retrograde.
Our analysis shows that narrow \sodium{} and \potassium{} absorption resonance lines are useful tracers of the atomic envelope of molecular clouds. In line with recent findings from giant molecular clouds, our results demonstrate that the velocity fields of the atomic and molecular gas are misaligned.  The angular momentum of a molecular cloud is not simply inherited from the rotating Galactic 
disk from which it formed but may be redistributed by cloud-cloud interactions.

\end{abstract}


\keywords{ISM: clouds – ISM: individual objects (Taurus) – ISM: kinematics and dynamics– (stars:) circumstellar matter – stars: formation –stars: kinematics}

\section{Introduction}
T Tauri stars (hereafter, TTs) are pre-main-sequence optically visible stars named after the brightest member of this class of objects in the Taurus Auriga molecular cloud \citep{joy42}. 
%Defining characteristics of TTs are optical variability and spectra with %strong %emission from hydrogen, helium, low excitation forbidden lines, and %metallic %emission resembling the solar chromosphere (e.g.\citealt{bertout89}). 
%
%Extensive multi wavelength observations have established that many of them %are surrounded by disks of gas and dust (e.g. \citealt{wc2011} for a %review). These circumstellar disks are the birth sites of planets and often %called protoplanetary disks. Hence, studying TTs is relevant to understand %both star and planet formation. 
%TTs belonging to the same star-forming region can present rather different %properties. 
Optical spectroscopy, and in particular the equivalent width of the H$\alpha$ line, is traditionally used to distinguish between two main subgroups: {\it classical} and {\it weak-line} TTs (CTTs and WTTs). CTTs are accreting disk gas as evinced by broad (FWHM$\sim$100\,km/s) permitted H and He lines tracing free falling gas onto the star, and by optical/UV emission in excess of the stellar photosphere probing the accretion shock at the stellar surface (e.g. \citealt{alcala2014}). On the contrary, WTTs have weaker emission lines with no evidence of ongoing accretion. Most WTTs also lack IR excess emission, hence are defined as Class~III objects, i.e. TTs with no circumstellar disks (but some WTTs have excess emission at longer wavelengths indicative of outer disks or dust belts, see e.g. \citealt{cieza2013}). CTTs have excess emission at multiple infrared wavelengths, typically starting at near-IR wavelengths, and are classified as Class~II objects (e.g. \citealt{espaillat2014}).

High-resolution (R$\sim$30,000) spectroscopy has been vital to identify and understand the complex processes associated with star and planet formation. In addition to accreting disk gas, CTTs eject 
%MH some
a fraction of this material which removes excess angular momentum. Forbidden emission lines blueshifted from the stellar velocity by $\sim$100\,km/s trace this phenomenon (see e.g. \citealt{ray2007} for a review). 
Forbidden lines also show a second low velocity component typically blueshifted by less than
10\,km/s that has been associated with a slow wind whose origin is still debated (e.g. \citealt{alexander2014} for a review and \citealt{natta2014}). Finally, permitted lines, including the H$\alpha$ line, often show complex morphologies indicating that they trace not only accretion but also outflowing material, such as winds originating within a few stellar radii (e.g. \citealt{mundt84}).
In this contribution we focus on the \sodium{} and \potassium{} resonance lines of TTs. 

While \potassium{} lines are generally not discussed in the context of TTs, several works have presented \sodium{} resonance line profiles of CTTs, focusing on the high accretors ($\dot{M} > 10^{-8}$M$_\sun$/yr) such as DL~Tau and DR~Tau. 
%MH:comment -- need to define "high accretor".  It would also be useful to give some examples or a prototype
The profiles are complex: in addition to emission, most spectra present absorption features that vary in width from being unresolved ($\le$10\,km/s) to $\sim$100\,km/s (e.g. \citealt{mundt84,edwards94}). \cite{ng1990} proposed that outflowing winds from the star can explain the P Cygni profile detected in about 20\% of the \sodium{} lines of CTTs, usually the most active ones. However, broad ($\sim$100\,km/s) \sodium{} emission profiles as observed in BP~Tau are most likely probing free falling accreting gas 
%MH as Balmer lines do
similar to the Balmer lines (e.g. \citealt{muzerolle98}). The most common feature in the \sodium{} lines of TTs is a narrow ($\le$10\,km/s) sharp absorption close to the stellar radial velocity tracing material physically in front of the star. The absorber has been usually associated with cold ISM gas (e.g. \citealt{mundt84}), although a circumstellar disk could not be ruled out with the available data (e.g. \citealt{appenzeller05}). A circumstellar origin would be especially interesting in the context of giant exoplanet atmospheres by providing a tracer of disk gas cooler than that being accreted onto the central star, gas that might be accreted onto the giant planets forming in the disk.

Recent transiting observations of hot Jupiters have found that the sodium-to-potassium abundance ratio in the atmosphere of these giant planets can be very different from the solar ratio (e.g. \citealt{sing2015}). 
% MH:comment: the following sentence is somewhat awkward 
Detailed theoretical models of planetary atmospheres demonstrate that photochemistry cannot explain such different ratios, and drastic changes to atmospheric temperature profiles or clouds are not supported by the data  (e.g. \citealt{lavvas2014}).
%MH added commas
% distinguished
% different 
%from those suggested by General Circulation Models or opacity from clouds,
% are not supported by the data  (e.g. \citealt{lavvas2014}). 
It is suggested that disk evolution might change the 
Na/K abundance ratios of the gas that giant planets accrete during formation. This hypothesis may be testable using optical spectroscopy of TTs in different evolutionary stages once the origin of their sodium and potassium lines is established.

This paper focuses on understanding the origin of the narrow and sharp absorption in the \sodium{} resonance profiles. Using high-resolution optical spectra for a sample of nearly 40 TTs in Taurus  spanning a range of evolutionary stages (Sect.~\ref{sect:obs}), we show that: a) \potassium{} profiles present the same sharp absorptions seen in \sodium{} profiles (Sect.~\ref{sect:analysis}); and b) The absorption does not arise from gas in a circumstellar disk, nor from ISM gas, but rather from atomic gas surrounding the Taurus molecular cloud (Sect.~\ref{sect:origin}). We also find that the atomic gas is more extended than the molecular gas, as traced by low J CO rotational lines, and that its rotation axis is almost anti-parallel to that of the Taurus molecular cloud. Our results, when combined with those on five giant molecular clouds in the Milky Way \citep{ib2011}, argue against the hypothesis that molecular clouds simply inherit the velocity field and angular momentum of the rotating galactic disk out of which they formed.

\section{Observations and Data Reduction}\label{sect:obs}
Observations were carried out with the Keck/HIRES spectrograph \citep{vogt94} in two separate campaigns, in 2006 through the program C170Hr (PI, L. Hillenbrand) and in 2012 through the program N107Hr (PI, I. Pascucci). In both campaigns we used the red cross-disperser with the C5 decker and a 1.1\arcsec $\times$7\arcsec{} slit. This configuration covers the 4,800-9,000\,\AA{} spectral range at a resolution of 37,500 as measured from the width of 4 Th/Ar lines near 5,240\,\AA{}\footnote{http://www2.keck.hawaii.edu/inst/hires/slitres.html}(but see below for our estimate of the spectral resolution). Exposure times for individual targets are summarized in Tables~\ref{tab:obs2006} and \ref{tab:obs2012}.

All science targets except TW~Hya are known members of the Taurus-Auriga star-forming region (age $\sim$1\,Myr, distance $\sim$140\,pc, e.g. the review by \citealt{kenyon08}). TW~Hya is a well characterized star that is closer to us and belongs to a loose association \citep{torres08}. We keep it in our analysis as comparing its \sodium{} and \potassium{} profiles to those of Taurus-Auriga members helps understanding the origin of the narrow and sharp absorptions in these resonance lines.
Targets from the 2006 campaign are mostly accreting classical T~Tauri stars with optically thick disks that extend inward close to the stellar surface (Class II SED), see Table~\ref{tab:psou2006}. In 2012 we targeted more evolved systems, with lower accretion rates on average, as well as non-accreting weak-line T~Tauri stars with transitional (small NIR but large MIR and FIR excess emission) or no disks (Class III SED), see Table~\ref{tab:psou2012}. Two sources, IP~Tau and UX~TauA, were observed in both campaigns. The central stars span a large range in spectral type, from M4 to K0, see Tables~\ref{tab:psou2006} and ~\ref{tab:psou2012} and \citet{hh14} for a careful and homogeneous re-classification of Taurus sources and of TW~Hya.
Along with science targets, we acquired a set of O/B stars to remove telluric lines and photospheric standards (both belonging to the Taurus-Auriga star-forming region as well as field dwarfs) to correct the science spectra from photospheric absorption features. 

The data reduction was carried out using the highly automated MAuna Kea Echelle Extraction (MAKEE) pipeline written by Tom Barlow\footnote{http://www2.keck.hawaii.edu/inst/common/makeewww/index.html}. The reduction steps included bias subtraction, flat-fielding, identification of the echelle orders on the images and extraction of the spectra. Wavelength calibration was done using spectra of ThAr calibration lamps mounted on HIRES, and acquired either right before or right after the target exposures. The reduction was performed in air wavelengths, using the keyword \lq \lq novac\rq \rq{} in the pipeline inputs. After spectral extraction, each of the individual 1-dimensional spectra was rebinned to a linear wavelength scale. The final products of the reduction are wavelength calibrated spectra to which heliocentric correction is also applied.

%MH -- there are many "we" cases in this paragraph.  I have rewritten some of these
In this work we focus on the two orders that cover the Na D resonance line at 5889.95\,\AA{}  and the \potassium{} line at 7698.96\,\AA\footnote{The weaker Na D resonance line at 5895.92\,\AA{} is too close to the edge of the order while 
the other \potassium{} resonance line at 7664.89\,\AA{} is not covered by our setting.}.
Of these two orders only that covering the \potassium{} line has strong telluric contamination, however no telluric 
%MH -- added line emission
line overlaps with the lines of interest. After telluric removal, the set of nearby photospheric 
standards are used  
%we used the set of nearby photospheric standards 
to subtract any photospheric absorption and identify those sources with Na and K emission. 
%%Note that the narrow absorption lines in the Na and K are already evident before 
%%photospheric subtraction in most sources. We made sure that our photospheric 
%%correction did not change the width and centroid of the narrow absorption lines which %%are the focus of this paper. 
The removal of photospheric lines follows standard procedures (e.g. \citealt{hartigan89}). In brief,
%MH we chose 
a photospheric standard with spectral type similar to that of the target star is chosen. When 
%MH needed we broadened 
necessary, the standard photospheric lines are broadened to match those of the target and the standard spectrum is shifted in wavelength to the target spectrum. The spectrum of the standard is also veiled to account for the extra 
emission known to fill in the absorption lines of young accreting stars. In practice we veiled the standard spectrum by identifying a flat continuum and a normalization factor that best reproduce the target spectrum in a specified wavelength interval outside of the lines of interest. Finally, we divided the target spectrum by the broadened, veiled, and shifted spectrum of the standard. The photospheric standard GL846 (M0 SpTy) provided a good correction for all stars except three later type M dwarfs (GH~Tau, V710~Tau, V1321~Tau) and eleven K dwarfs (CI~Tau, CW~Tau, GK~Tau, GM~Aur, HBC~427, HN~Tau, HQ~Tau, IT~Tau, UX~Tau, V773~Tau, V1348~Tau). For the group of late M dwarfs
%MH  we used 
the photospheric standard GL15 (M2 SpTy) was used while for the K dwarfs, the standard HR~8832 (K3 SpTy) was applied. The approach described above produces residual line profiles. Care should be taken to interpret these profiles, especially weak and/or narrow excess emission, because \sodium{} and \potassium{} photospheric lines are deep and gravity 
%MH sensitivity
sensitive. However, 
%MH we must use
 nearby/older dwarfs are necessary to remove the photosphere because even WTTs in Taurus present the deep \sodium{} and \potassium{} absorption lines we want to investigate. To illustrate the difficulty in creating residual profiles 
we show in 
Fig.~\ref{fig:sample_resid} the original spectra, the veiled/scaled and broadened photosphere, and the residual spectra for three stars with SpTy from early K (V773~Tau) to middle M (GH~Tau) and covering both CTTs and WTTs. Residual profiles are typically less reliable for WTTs than CTTs.
%MH  broke this up into 2 sentences
Further discussion of these profiles is provided in Sect.~\ref{sect:residual_profiles}.  

We also independently measured the spectral resolution achieved by HIRES in our observational setting using the strongest telluric lines in the order covering the \potassium{} resonance transition. The distribution of FWHMs for nearly 90 telluric lines from spectra obtained in the 2006 and 2012 campaigns is shown in Fig.~\ref{fig:fwhms}. A gaussian fit to the distribution provides a mean of 6.6\,km/s and a standard deviation of 0.5\,km/s. The spectral resolution of $\sim$45,000 that we obtain is slightly better than that reported in the Keck webpage but consistent with that value within three times the standard deviation.

\section{Analysis}\label{sect:analysis}
As mentioned in the Introduction, sharp absorptions in the \sodium{} (D) resonance lines of TTS were noted early on. In 27 out of our 39 targets we find that a similarly sharp absorption is also present in the \potassium{} resonance line at 7698.96\,\AA{} (see Figs.~\ref{fig:NaKLi_ctts1},~\ref{fig:NaKLi_ctts2},~\ref{fig:NaKLi_ctts3},~\ref{fig:NaKLi_wtts}, and \ref{fig:NaKLi_multi}). 
We take advantage of the high spectral resolution of HIRES and our large sample of stars covering different evolutionary stages to understand the origin of these sharp features.
For each star we calculate the centroid of the absorption and the FWHM and EW of the line (see Tables~\ref{tab:RV2006} and \ref{tab:RV2012}). We then compare these values to stellar radial velocities and the velocity and width of low J CO emission lines tracing the cold ($\sim$15\,K) molecular gas in the Taurus star-forming region.

To compute stellar radial velocities we measure the centroid of two prominent photospheric lines that fall in different spectral orders: the Li doublet at a mean wavelength of 6707.83\,\AA{} (our spectral resolution is not sufficient to resolve the doublet) and the Ca line at 6439.07\,\AA{}. We apply a Monte Carlo approach to evaluate the uncertainty in the peak centroid by adding a normally distributed noise to each spectrum and computing the radial velocity thousand times for the same star. This uncertainty is typically $\sim$0.2\,km/s except for HN~Tau and V410~Tau for which RV uncertainties are $\sim$3\,km/s, see also Tables~\ref{tab:RV2006} and \ref{tab:RV2012}.
The spectrum of HN~Tau has a rather low S/N and the photospheric lines are broad and shallow. Although our radial velocity for HN~Tau of 13.8\,km/s is uncertain, it is consistent with the median heliocentric radial velocity of Taurus members which is 16.2\,km/s \citep{n12}.  V410~Tau has a stellar companion (see Table~\ref{tab:psou2012}) and its photospheric lines present multiple components, thus our fit is sensitive to the wavelength region chosen to fit the line. {Two targets HBC~427 and V773~Tau are well known spectroscopic binaries. For these sources we report the barycenter velocity of the systems calculated from multi-epoch spectroscopic campaigns (see Table~\ref{tab:RV2012}).
The standard deviation between the RVs obtained from the Li and Ca lines is $\sim$0.8\,km/s. 
Most of the stellar RVs we report in Tables~\ref{tab:RV2006} and \ref{tab:RV2012} are from the Li line except in four instances where this photospheric line appears slightly asymmetric or flat making it difficult to have an accurate centroid. In these cases, clearly marked in the Tables, we use the RV computed from the Ca line at 6439.07\,\AA{}. 
Comparison of our RVs to literature radial velocities yields differences (excluding the sources with large uncertainties mentioned above) of at most 4\,km/s with the standard deviation being $\sim$1\,km/s. Thus, unless specified in Tables~\ref{tab:RV2006} and \ref{tab:RV2012} we assign a one sigma uncertainty of 1\,km/s to the stellar radial velocities reported in this work. 

To measure the peak centroid of the molecular gas at the location of our Keck targets we have used two CO maps of Taurus. One map was obtained with the FCRAO 14\,m telescope in the $^{12}$CO and $^{13}$CO $J=1-0$ transitions (beam 45$\arcsec$, \citealt{narayanan08}) while the other is a $^{12}$CO $J=1-0$ map acquired with the CfA 1.2\,m telescope (beam 8.4$\arcmin$, \citealt{dame01}). Ten of our Keck targets\footnote{The targets that are not covered by the FCRAO map are: DR~Tau, HN~Tau, UX~Tau, DM~Tau, GM~Aur, HBC~427, V710~Tau, V836~Tau, V1321~Tau, V1348~Tau} lie outside the higher spatial resolution FCRAO map but are covered by the coarser CfA $^{12}$CO map. 
For all targets we extract the $^{12}$CO and the $^{13}$CO spectra at the source location and compute the centroid velocity over a $\sim$6\,km/s window centered around the Taurus mean velocity, $\sim$6\,km/s LSR (e.g. \citealt{goldsmith2008}), the uncertainty in the centroid, and the signal-to-noise of the CO spectrum.
The $^{12}$CO and $^{13}$CO centroids from the FCRAO map are very similar, their standard deviation being only 0.2\,km/s, but the $^{12}$CO profiles are broader (typical full width at half maximum of $\sim 1.5$\,km/s, see also Goldsmith et al.~2008). 
%We find that the median uncertainty of the FCRAO centroids is 0.05\,km/s while %that of the CfA centroids is larger, 0.2\,km/s. 
For spectra with a signal-to-noise greater than 5 and sources in common in the FCRAO and CfA surveys we find that the standard deviation in the peak centroids is $\sim$0.4\,km/s and there are no cases for which the absolute difference in the centroids is larger than 1\,km/s. 
Given that the uncertainty in the RVs obtained from the optical spectra is $\sim$1\,km/s, we have opted to provide the peak centroids and the signal-to-noise of the CO spectra extracted from the CfA Taurus map which covers all of our targets (see Tables~\ref{tab:RV2006} and \ref{tab:RV2012}).

%The slit we employ is only 1.1\arcsec $\times$7\arcsec{} drastically %reducing spatial contamination but we may still be detecting multiple %atomic clouds in the line of sight of our sources due to our much lower %spectral resolution. This is obvious in the case of 
%HQ~Tau and CI~Tau where two velocity components are detected in the %%\potassium{} absorption profile, but not in the \sodium{} absorption. As %discussed previously, we think that the resolved widths of many \sodium{} %lines are due to a combination of line saturation and multiple cloud %components and do not reflect more turbulence or higher temperatures in the %sodium gas than in the potassium. 

\subsection{Residual \sodium{} and \potassium{} profiles}\label{sect:residual_profiles}
A comparison of residual \sodium{} and \potassium{} resonance profiles and Li photospheric lines is shown in Figs.~\ref{fig:NaKLi_ctts1},~\ref{fig:NaKLi_ctts2},~\ref{fig:NaKLi_ctts3},~\ref{fig:NaKLi_wtts}, and \ref{fig:NaKLi_multi}. Because \potassium{} absorption lines are on average a factor of $\sim$2 weaker than 
\sodium{} lines we have scaled these profiles when needed, scaling factors are provided in the figure captions. In all objects except TW~Hya and perhaps V1348~Tau (Fig.~\ref{fig:NaKLi_wtts}) it is possible to identify one absorption feature in the \sodium{} profile that is narrower than the photospheric Li line and it is close to, but not always coincident with, the stellar RV. Similar narrow absorptions are present in the \potassium{} profiles of about two thirds of the targets. 
%The velocity of the \sodium{} and \potassium{} absorption features is close to but not always coincident with the %stellar radial velocity as measured from the Li lines.

Five sources (DO~Tau, CI~Tau, HQ~Tau, IT~Tau, and V836~Tau) present one additional narrow absorption that is more offset from the stellar RV (see Fig.~\ref{fig:hqtau-citau}). Except for CI~Tau and HQ~Tau this extra absorption is only clearly detected in the \sodium{} profiles.
For IT~Tau the additional absorption is at v$_{\rm helio} \sim$-2\,km/s, about 18\,km/s blueshifted from the stellar RV, while for DO~Tau and V836~Tau it is at $\sim$7.3\,km/s, 10 and 12 km/s blueshifted from the stellar RV. CI~Tau and HQ~Tau are at only 0.5 degree projected distance from each other and both present extra absorption at higher velocities, specifically at 23.4 and 27.3\,km/s respectively. This implies that the absorption is redshifted by 4.3 and 10.2\,km/s from the stellar RV. 
%Because these extra narrow absorption are rare, mostly present only in the \sodium{} %profiles, and weaker than those close to the stellar RV, we will not discuss them %further in the text.

As previously reported in the literature, \sodium{} profiles of TTs present a range of morphologies. More than half of our sample stars (23/39) show \sodium{} emission with the highest accretors having FWHMs clearly broader than photospheric lines (see e.g. CY~Tau and DO~Tau in Fig.~\ref{fig:NaKLi_ctts1}). Magnetospheric infall is the most likely origin of this emission but stellar winds may also contribute. Conversely, the \sodium{} emission of WTTs is similar in width to (or narrower than) photospheric lines (see Fig.~\ref{fig:NaKLi_wtts}).  While some of this excess emission may result from an improper photospheric subtraction (see Sect.~\ref{sect:obs} for details), the stellar chromosphere may also contribute. Broad absorptions in the \sodium{} profiles that are clearly outside the range of photospheric velocities are detected toward five stars (DF~Tau, DG~Ta, DL~Tau, DO~Tau, and DR~Tau, see Fig.~\ref{fig:naoi_jets}). The restricted velocity range adopted in Figs.~\ref{fig:NaKLi_ctts1},~\ref{fig:NaKLi_ctts2}, and \ref{fig:NaKLi_ctts3} only shows the blueshifted component of the broad absorption, which is associated with ejected material. However, in all cases there is also redshifted Na absorption that traces infalling disk gas onto the star (see Appendix~\ref{sub:broadNa}). We note that for all of the sources in our sample, \potassium{} profiles are less complex than \sodium{} profiles. This is evident both in the sample of high accretors ($\dot{M} \ge 10^{-8}$\,M$_\sun$/yr), only DG~Tau and DO~Tau have a broad blueshifted \potassium{} absorption beyond the photospheric line, and in the sample of low/no accretors, see e.g. the presence of only one absorption line in the \potassium{} residual profiles of IT~Tau and V773~Tau. 
In the next sections, we will discuss the origin of the narrow Na and K absorptions, excluding the second components seen in five stars with velocities differing significantly from the stellar velocity.

\subsection{Empirical trends}\label{sect:narrowNa}

In this section we present empirical trends between the narrow and sharp \sodium{} and \potassium{} absorption features, extinction along the line of sight, stellar radial velocities, and the velocity of the cold molecular gas in Taurus.

As mentioned earlier a deep narrow \sodium{} absorption is present in all spectra except those of V1348~Tau and TW~Hya while \potassium{} absorption is present in two thirds of the spectra. 
With perhaps the exception of V836~Tau, the \sodium{} lines appear highly saturated, hence their EWs cannot be used to infer the column density of the absorbing material. Saturation is also common in the \potassium{} lines when detected.
For both transitions line FWHMs span a larger range of values (by a factor of $\sim$3) than RVs and are not correlated with them.
When absorptions are present both in the \sodium{} and \potassium{} profiles, their RVs are the same within our $\sim$1\,km/s velocity accuracy, the median absolute value of their difference being 0.6\,km/s. This is illustrated in the upper left panel of Fig.~\ref{fig:NaK_comp} where a linear fit to the centroids of the \sodium{} and \potassium{} absorptions results in a slope very close to unity.  
%One outlier is the source HQ~Tau for which the \potassium{} line is redshifted by 3.5\,km/s with respect to the \sodium{} line %when only one gaussian is used to fit the profile. However, the \potassium{} profile from HQ~Tau, as well as that of the nearby %source CI~Tau, clearly shows two components, one of them coincident with the broad \sodium{} absorption line (see Fig.~%%\ref{fig:hqtau-citau}). 
Although the EWs of the absorption lines are more sensitive to the photospheric subtraction than peak centroids, hence less accurate, the \sodium{} and \potassium{} EWs are correlated. The Kendall’s $\tau$ test (see e.g. \citealt{press93}) on the subset of sources with detections in both lines returns $\tau=0.4$ and a probability of $5\times 10^{-4}$ that the variables are uncorrelated. These results demonstrate that there is a significant positive correlation between the \sodium{} and \potassium{} EWs, with the \potassium{} absorption, when detected, being a factor of $\sim$2 lower than the \sodium{} absorption (Fig.~\ref{fig:NaK_comp}, upper right panel). The fact that Na and K sharp absorptions occur at the same heliocentric radial velocity {\it and} their EWs are positively correlated strongly suggests that the two atomic transitions trace the same gas. One notable difference in their absorption profiles is the line width: many \sodium{} absorption lines are spectrally resolved (the mean FWHM is 8.2\,km/s) while most \potassium{} lines are unresolved with a mean FWHM of 6.3\,km/s, the same as that measured for telluric lines. We will discuss this further in Sect.~\ref{sect:origin}.

%{\bf TO ADD: Say that there is no correlation between the FWHM of the lines and the peak velocity also there %is no correlation between Av and the peak centroid}

 We also test if there is any correlation between the optical extinction along the line of sight (A$_{\rm V}$) and the \sodium{} and \potassium{} EWs, line-to-continuum ratios\footnote{The line-to-continuum ratio is defined here as the depth of the absorption line with respect to the local continuum.}, and RVs. It is known that extinctions are rather uncertain and depend on the specific technique and wavelength range used to derive them (see e.g. discussion in \citealt{edwards2013} and \citealt{mc2014}). Recently, \citet{hh14} have re-computed spectral types, extinctions, and accretion in a homogeneous way for over 280 nearby TTs using an improved technique that fits simultaneously these three parameters to flux calibrated optical spectra. All but four of our sources overlap with their sample, hence we use this homogenous set of A$_{\rm V}$ for our analysis. In spite of a relatively large scatter, which may be due to our use of photospheric templates to compute the residual profiles of these gravity-sensitive lines, the Kendall $\tau$ test suggests a moderate positive correlation between the line EWs and A$_{\rm V}$, $\tau=0.4$ and a lower probability that the variables are uncorrelated for Na ($5\times 10^{-3}$) than for K ($7\times 10^{-2}$), see also Fig.~\ref{fig:NaK_comp} lower left panel.
The \sodium{} line-to-continuum ratios are less correlated with A$_{\rm V}$ but a modest correlation remains between the \potassium{} line-to-continuum ratios and A$_{\rm V}$, $\tau=0.4$ and probability of being uncorrelated of $2\times 10^{-2}$ (Fig.~\ref{fig:NaK_comp} lower right panel). Finally, the RVs of the \sodium{} lines are not correlated with A$_{\rm V}$ ($\tau=0.01$ and probability of 0.9). These results hint to a possible relation between the gas probed via the Na and K absorption lines and the location of our targets in the Taurus molecular cloud.
%{\bf TO ADD: MAYBE ADD HERE THAT THERE IS FOR INSTANCE NOT A GOOD CORRELATION ETWEEN THE AMOUNT OF NH TWARD %THE STAR AND THE AV AS SHOWN IN MC JUNKIN SO IT IS NOT CLEAR THAT WE ARE REALLY PROBING WITH AV THE JUST THE %MATERIAL BEYOND THE STAR, IN GENREAL AV ARE LARGER BUT NOT CORRELATED WITH THE NH FROM LYALPHA.}

We now turn to the comparison between the RVs of the Na and K absorption features and the stellar radial velocities provided in Tables~\ref{tab:RV2006} and \ref{tab:RV2012}. A histogram of the difference between these quantities is shown in Fig.~\ref{fig:Histo_Li_Na_CO}. According to the non parametric Kolmogorov-Smirnov (hereafter, K-S) statistics, there is a $\sim$12\% probability that the two velocity distributions are drawn from the same parent population, in other words they are statistically indistinguishable. Indeed,  
for many sources the Na and Li RVs  are the same given our velocity uncertainty. However, there are sources for which the absorption lines are clearly redshifted (e.g. VYTau, DM~Tau, UXTau~A) or blueshifted (e.g. BP~Tau, CY~Tau, V819~Tau) with respect to the stellar radial velocity, see Figs.~\ref{fig:NaKLi_ctts1},~\ref{fig:NaKLi_wtts}, and~\ref{fig:NaKLi_multi}. This group comprises CTTs as well as WTTs, Class II and III SEDs, as well as TDs. One thing in common to these sources with deviant velocities is that they lie at the edge of the Taurus molecular cloud, see  Sect.~\ref{sect:origin}. 
Thus, atomic Na and K gas is not always associated with the immediate circumstellar environment.
% http://www.haystack.mit.edu/edu/undergrad/srt/SRT%20Projects/rotation.html

In further comparing the stellar and \sodium{} RVs with the velocity of the molecular gas we restrict ourselves to CO spectra with a signal-to-noise ratio greater than 5, thus excluding HN~Tau, DM~Tau, HBC~427, GM~Aur, VY~Tau, V1321~Tau, V1348~Tau. As already reported in the literature (e.g. \citealt{hartmann86}), we do not find any statistically significant difference  between the stellar RVs and the RVs of the molecular CO gas at the source location (K-S probability $\sim$12\%, see also Fig.~\ref{fig:Histo_Li_Na_CO}). This is expected because stars form in molecular clouds, thus inherit the cloud velocity field, and clouds are not dispersed in $\sim$1-2\,Myr. Interestingly, the distribution of RVs for the atomic and molecular components are statistically different, with a K-S probability of only 0.6\% that they are drawn from the same parent population, see also Fig.~\ref{fig:Histo_Li_Na_CO}. This strongly suggests that the molecular and atomic gas are not co-located. Sources with low signal-to-noise CO detections but strong \sodium{} absorption, such as VY~Tau and DM~Tau, further strengthens this statement (see Fig.~\ref{fig:NaKLi_wtts}) and suggests that the atomic gas is more extended than the molecular gas.  We also note that the FWHMs of the $^{12}$CO lines are much narrower than those of the \sodium{} lines and would all appear unresolved at the resolution of our optical spectra. However, this difference in FWHMs may be due to a combination of saturation of the \sodium{} lines and the presence of other atomic clouds along the line of sight, hence it does not further constrain the spatial location of the molecular and atomic gas.

To summarize, we find that: i) \sodium{} and \potassium{} absorption lines trace the same atomic gas but \sodium{} profiles are often spectrally resolved while \potassium{} lines are not; ii) There is a modest correlation between the EW of these absorption features and A$_{\rm V}$; 
%iii) Some of the \sodium{} and \potassium{} absorption lines have RVs substantially different %from the stellar radial velocities; 
and iii) Although there is no statistical difference between the stellar RV distribution and either the \sodium{} or CO radial velocities, the atomic and
molecular gas are unlikely to be drawn from the same parent population.

%{\bf The distribution of CO and \sodium{} RVs are statistically different while the CO and %stellar RVs are not}.

\section{Origin of the narrow absorption in the \sodium{} and \potassium{} profiles}\label{sect:origin}
%MH -- I don't think you need these underlined subheadings and think it would be better to remove each.
%MH {\bf \underline{Circumstellar disk hypothesis.}} 
Previous studies 
%MH
of Na absorption lines
 could not distinguish between a disk or an interstellar/cloud origin because they investigated fewer stars, mostly CTTs with Class~II SEDs, at a lower spectral resolution than achieved here.  
%MH Need a reference here 
By extending our sample to include WTTs and Class III SEDs and with a precision in radial velocity of $\sim$1\,km/s,  we can exclude that the sharp and narrow \sodium{} and \potassium{} absorption lines trace circumstellar disk gas. There are three main observables pointing against a disk origin. First, strong absorption 
%MH
features are
%is 
also detected toward objects that are not surrounded by disks (e.g. Anon1) or have only tenuous dust disks more akin to debris disks (e.g. V1321~Tau). Second, if the absorbing material were disk gas there should be only one narrow absorption feature toward single stars and its velocity should be coincident with the stellar radial velocity. As discussed in the previous sub-section this is not the case for several systems. 
%MH next sentence is unncessary
%Excluding the disk hypothesis leaves open the possibility for an interstellar/intracloud origin which we discuss next.
Third, as will be shown later in this section, there is a spatial gradient in the RV of the \sodium{} lines
indicating a location in the distributed gas rather than the immediate circumstellar environment.

%MH {\bf \underline{Interstellar hypothesis.}} 
%MH labeling local bubble material as interstellar gas is confusing.  We should refer to this simply as the local bubble 
\citet{rl08} constructed an empirical dynamical model of the local ISM (inside the local bubble, i.e. within about 100\,pc) 
using UV HST spectroscopy toward 157 sight lines.
While their model relies on the more numerous \caii{} absorption lines,
they find that all \sodium{} absorption components have companion \caii{}
absorption at the same velocity. This indicates that the gas detected by the
\caii{} absorption is physically associated with that detected with the
\sodium{} line.
%About one-half of their \sodium{} detections are toward stars 
%within 50\,pc suggesting that \sodium{} absorption is not 
%dominated by the cold gas at the edge of the local bubble. 
Nine of their stars are in the direction of the Auriga Cloud.
We use the cloud heliocentric velocity vectors they derive 
(their Table~16) to compute the RV of the local bubble gas at the location 
of each of our targets. With a mean predicted RV of 20\,km/s and minimum and maximum of 19\,km/s and 23\,km/s, the gas in the local bubble cannot be responsible for the 
average \sodium{} RV of 17.6\,km/s and the larger range of velocities 
(from $\sim$13 to 23\,km/s) we find in our sample. 
Only 13 out of 38 unique Keck targets have RVs consistent with those predicted 
by the local bubble model of \citet{rl08} and the K-S test gives a probability of 
only $3\times 10^{-8}$ that the two velocity distributions are drawn from the 
same parent population. In addition, of the nine stars in the direction of
the Auriga Cloud, only one of them has a detected \sodium{} absorption component 
with an EW of only $\sim$3\,m\AA{} \citep{far2015}, much smaller than the absorption 
we measure toward our TTs.
We conclude that the \sodium{} and \potassium{} absorption features we detect are 
not associated with local bubble gas. 
This conclusion is also consistent with the lack of absorption in the \sodium{} and \potassium{} profiles of TW~Hya, which is at only $\sim$50\,pc from the Sun \citep{mamajek05}, albeit in a different direction from Taurus. 

%MH new paragraph
The distribution of interstellar gas and its kinematic beyond the local bubble is less well 
%MH determined.
established. 
\cite{wh2001} obtained high-resolution (FWHM$\sim$0.4-1.8\,km/s) spectra of \potassium{} absorption toward 54 Galactic OBA stars. One of them, HD~27778 (62~Tau), is in the direction of the Taurus star-forming region at a projected separation of $\sim$1.5$^\circ$ from DF~Tau and $\sim$1.9$^\circ$ from DG~Tau and FZ~Tau. At a distance of 225\,pc, as measured by  {\it Hipparcos}  (ESA 1997), HD~27778 is not a Taurus member but likely belongs to the older Cas-Tau OB association located behind the Taurus molecular cloud (e.g. \citealt{mooley13}). The \potassium{} profile obtained by \cite{wh2001} shows two deep and spectrally resolved absorptions at $\sim$14.8\,km/s (consistent with the DF~Tau and DG~Tau \sodium{} and \potassium{} velocities) and $\sim$18.5\,km/s (slightly lower than the FZ~Tau velocities). The \sodium{} absorption profile along the same line of sight is highly saturated (see their Fig.~2), because sodium is $\sim$15 times more abundant than potassium. This means that the identification of individual clouds is more challenging using \sodium{} lines even at high spectral resolution. Saturation and blending of clouds at similar velocities might explain why many \sodium{} absorption lines toward the Taurus TTs are resolved at our spectral resolution, $\sim$4 times lower than that of \cite{wh2001}, while most \potassium{} lines remain unresolved. As pointed out in Sect.~\ref{sect:narrowNa}, this is clearly the case for HQ~Tau and CI~Tau where two cloud components are detected in the \potassium{} but remain blended in the \sodium{} profiles.
\cite{wh2001} model the \potassium{} profile observed toward HD~27778 with six
atomic clouds in the foregrounds, four of them have RVs within 2\,km/s of the
median stellar RVs of Taurus members. For these clouds they estimate
\potassium{} column densities respectively between 1.4 and 1.9$\times 10^{11}$\,cm$^{-2}$, which correspond to an average $\sim 6 \times 10^{20}$\,cm$^{-2}$ hydrogen column given their empirical relation between K and H column densities. As a comparison \citet{mc2014} find H column densities of  $\sim 2.5 -10 \times 10^{20}$\,cm$^{-2}$ toward 12 TTs in Taurus from neutral hydrogen absorption against broad Ly$\alpha$ emission profiles.
These column densities are consistent with each other, about two orders of magnitude larger than the H column densities measured inside the local bubble \citep{rl08}, but rather typical to ISM clouds \citep{wh2001}.

%The similarity in column densities suggests that we may be probing atomic gas associated %with the Taurus molecular cloud rather than ISM gas beyond the local bubble.

%MH {\bf \underline{Intracloud hypothesis.}} 
%MH Here, we test if 
Do the \sodium{} and \potassium{} narrow absorption features trace atomic gas associated with the Taurus molecular cloud or rather ISM gas beyond the local bubble? To answer this question we plot the location of our TTs on the sky and color code the radial velocity of the absorbing material toward them (see Fig.~\ref{fig:colormap})\footnote{We use the RV of the \sodium{} lines because we have more detections in this line than in the \potassium{} line and have already demonstrated that when both lines are detected there is a good coincidence between their RVs.}. The figure reveals that nearby sources have similar velocities and a clear spatial gradient in RVs with lower RVs NW and higher RVs SE. Several authors have argued that linear velocity gradients in Giant Molecular Clouds are the result of large-scale cloud rotation (e.g. \citealt{blitz93}). Following common practice and assuming solid-body rotation (e.g. \citealt{goodman93}), we fit a plane to identify the best-fit linear velocity gradient and measure its magnitude ($\Omega$) and direction ($\theta$), i.e. the direction of increasing velocity measured east of north. The rotation axis of the cloud is then $\theta + 90^\circ$. With this approach we find that $\theta \sim 125^\circ$ (P.A. of the rotation axis 215$^\circ$)\footnote{These values are obtained including V1348~Tau. If we exclude this source we find P.A.=214$^\circ$ and $\Omega$=0.11\,km/s/pc basically the same as those reported in the text.} which is the same as Galactic rotation, 
% MH The more conventional unit for Omega is km/s pc-1 so omega=0.113 km/s/pc
$\Omega \sim 0.11$\,km/s/pc, 
and the mean LSR cloud velocity is 7.2\,km/s. To demonstrate that a plane is a good fit to the velocity map, we show in Fig.~\ref{fig:position_velocity} a position-velocity diagram passing through the center of the CO map, where the position is the offset of each star along the direction perpendicular to the rotation axis. As expected for gas rotating as a solid-body, we see a linear trend in the velocity as a function of offset from the rotation axis. Thus, it appears that the atomic gas associated with the Taurus molecular cloud rotates in the same direction of Galactic rotation \citep{kd1985}. Evidence for global rotation of the Taurus complex has been recently reported by Rivera et al. (2015) from the RVs and proper motion measurements for 7 well known Taurus members. These authors find velocity gradients of $\sim$0.1\,km/s/pc and positive direction of rotation with respect to Galactic rotation as we find from the study of the \sodium{} lines.

Velocity gradients in Taurus have been also measured using other tracers and it has been long known that the Taurus velocity field is rather complex even on large $\sim$10\,pc scales (e.g. \citealt{shuter1987}). 
Here, we are particularly interested in comparing the atomic and the molecular components. Fig.~\ref{fig:colormap13CO} shows the \sodium{} velocities (as stars) superimposed on the higher spatial resolution $^{13}$CO FCRAO map\footnote{We use the FCRAO map because it has higher spatial resolution and only four sources are detected beyond this map in the coarser CfA survey.}.
The $^{13}$CO gas shows a SE-NW gradient with a rotation axis of 25$^\circ$, $\Omega \sim 0.25$\,km/s/pc, and a systemic LSR velocity of 6.4\,km/s \citep{shuter1987}. The $^{13}$CO rotation is almost anti-parallel to that of the \sodium{} gas and retrograde with respect to differential galactic rotation. 
%Fig.~\ref{fig:colormap13CO} illustrates the different velocity fields that we are %about to discuss. Here, velocities are converted to LSR to better compare with %already published work.
%The $^{13}$CO gas shows a SE-NW gradient with a rotation axis of 25$^\circ$, 
% \Omega \sim 0.25$\,km/s/pc, and a systemic LSR velocity of 6.4\,km/s. 
% The HI 21\,cm absorption, whose column density distribution is similar to that %of the $^{13}$CO, has a systemic velocity $\sim$3\,km/s lower than the  $^{13}$CO %gas and a NE-SW gradient suggesting a rotation axis for the cold atomic gas of
% P.A.=297$^\circ$, almost perpendicular to that of the molecular gas
%MH
% \citep{shuter1987}. 
%MH \citet{shuter1987}
%MH  the following sentence is rather speculative on the part of Shuter etal. 
%These authors  interpret the properties of the HI absorbing gas as due to %velocity waves coupled with magnetic field line vibrations. 
%MH  Not convinced that HI emission tracks CO very well.  Certainly not from these 1987 maps and even 
% with more recent 21cm maps (see Krco  )  The larger FWHM arise from the warm (T=8000 K) component and may not even be physically linked with Taurus
%{\bf COMMENT/TODO: MH is not convinced that the 21cm HI emission tracks better %the CO gas, double ck literature including Kro et al. 2010.}
Interestingly, \citet{balle99} find that the HI 21\,cm emission has also a velocity offsets of a few km/s and a larger velocity dispersion ($\sim$5-10\,km/s LSR) than the molecular CO gas, resembling the one we see in the \sodium{} and \potassium{} absorption lines. Position velocity diagrams along some directions show a different axis of rotation from the CO gas (e.g. Fig. 3a of \citealt{balle99}) but the magnitude and direction of rotation of the HI emitting gas has not been quantified. 
As a note, the magnetic field in Taurus has a mean P.A. of 25$^\circ$(=205$^\circ$), the same as the CO and \sodium{} rotation axes. This is consistent with the classic theory of isolated low-mass star formation whereby cores contract more along magnetic fields, resulting in a rotating flattened morphology perpendicular to the field axis (e.g. \citealt{shu1987}). What is surprising is that the atomic and molecular gas rotate in almost opposite directions. What is the relation between the atomic and molecular gas?

%MH If you want to include a physical mechanism, then magnetic braking would produce misaligned/orthogonal 
% rotational axes between cloud and extended envelope

%Following the same color scheme as \citet{goldsmith2008}, we show  a %color-coded image of the $^{13}$CO integrated intensity\footnote{The %large scale velocity gradient measured using $^{12}$CO is the same as %$^{13}$CO. We prefer to show the  $^{13}$CO intensity map because %cloud structures are more apparent, see \citep{goldsmith2008} for a %comparison of the $^{12}$CO and $^{13}$CO maps} and plot on top of it %our sources color coded based on their \sodium{} RVs.

%%http://scitechdaily.com/herschel-discovers-a-newfound-reservoir-of-stellar-fuel/
%% NOTE: That is, east is the direction of increasing right ascension.
%%% RELATION BETWEEN AV AND NA EW, NOT SURE WE NEED THIS!
%4. Although our EW measurements are rather uncertain we see that there is a correlation between the %extinction measured toward individual objects (as done in HH2014) and the EW of the sodium and a linear %relation between the Na and K EWs as expected from the coincidence in velocity of the clouds. I NEED TO %EXPLAIN WHY THE LINE WIDTH OF K AND NA IS DIFFERENT. It seems like the reason is that Na samples more %than one component while K, being less abundant tan Na by a factor of 15, only sample the higher NH %cloud
%
%% NOTE for myself : Ten of our Keck sources are outside the area mapped by  Narayanan et al.~(2008), %these sources are DR~Tau, HN~Tau, UX~TauA, DM~Tau, GM~Aur, HBC~427, V710~Tau, V836~Tau, V1321~Tau, and %V1348~Tau

The larger velocity dispersion in the \sodium{} and \potassium{} absorptions and in the HI 21\,cm emission compared to the CO peak velocities suggests that the atomic gas is more widely distributed. Indeed, these atomic tracers are detected even at locations where molecular CO gas is not detected. Thus, they appear to trace the atomic gas that is  the leftover of star-formation. In addition, the mere detection of alkali metals in the gas phase implies that there are regions warmer than those traced by CO, regions that are not shielded from the ambient FUV radiation, else Na and K would be depleted onto the surface of dust grains (e.g. \citealt{barlow1978}). We can further place an upper limit on the gas temperature that would result in thermal broadening (for the moment we neglect turbulence) by assuming that each atomic cloud is spectrally unresolved at our 6.6\,km/s resolution (such as for instance toward BP~Tau or V819~Tau). Sodium gives a more stringent constraint than potassium, given its lower atomic mass, but not a tight one: the gas temperature should be $\le$ 2,000\,K for sodium lines to be unresolved at our spectral resolution. 
A much tighter constraint comes from the higher resolution spectra of \cite{wh2001}. These authors measure \potassium{} line-width parameters (b$\sim$1.667 FWHM) for over 100 ISM clouds, including those in the direction of HD~27778, and find a distribution peaking at $\sim$0.7\,km/s. Assuming an isotropic turbulence of $\sim$0.5\,km/s, which seems to be common to many lines of sight, implies an atomic gas temperature of $\sim$100\,K. Gas as cold as CO ($\sim$15\,K) would result in b-values much smaller than typically observed.
%A much tighter constraint comes from the higher resolution spectra of \cite{wh2001}. Using %the \potassium{} resonance lines they measure FWHMs of $\sim$0.7\,km/s for the clouds in the %line of sight of HD~27778, which sample the Taurus atomic gas. These FWHMs imply a gas %temperature of at most 400\,K when turbulence is neglected.  Adding a subsonic turbulence of %$\sim$0.5\,km/s, which seems to be common to many lines of sight in the sample of %%\cite{wh2001}, further reduces the atomic gas temperature to $\sim$200\,K. 
These results suggest that the gas associated with the alkali metals is much warmer than the molecular gas. In models of photodissociation regions, temperatures of several hundred kelvin also trace the transition zone where ionized carbon is converted into atomic carbon. This zone is well outside the cold and FUV-shielded core where most of the hydrogen is in H$_2$ and most of the carbon in CO (e.g. \citealt{roellig2007}). 
Recently, Herschel/HIFI observations demonstrated that a significant fraction, between 20 and 75\%, of the ISM gas that is undetected in HI and CO is bright in the [C~II] FIR line at 158\,\micron{} \citep{langer2014}. \citet{orr2014} compared [C~I], [C~II], and CO emission lines toward a linear edge region in the Taurus molecular cloud. While the strength of the [C~I] lines is well correlated with that of CO lines (and its $^{13}$CO isotoplogue), the [C~II] lines are only detected  (at a low S/N) in two locations outside of the region where most of the molecular emission occurs (see their Fig.~1). A comparison of [C~II] peak centroids and \sodium{} and \potassium{} absorption toward the same line of sight would be very valuable to test if these transitions probe similar or larger regions.

To summarize, the 
%MH data available so far 
available data suggest that the \sodium{} and \potassium{} absorption lines trace the warm ($\sim$100\,K) atomic envelope of the cold molecular CO gas in Taurus. Its extent may be comparable to that of the HI 21\,cm and/or the [C~II] FIR emission. The finding that the rotation axes of the atomic and molecular gas are not aligned is not unique to the Taurus region.  \citet{ib2011} compared the velocity gradient P.A. of the molecular (CO) and atomic (HI 21\,cm emission) gas of five giant molecular clouds in the Milky Way. Except for Orion A, they find that
%MH their position angles 
the positional angles of the molecular and atomic morphological axes 
are widely separated, by as much as 130$^\circ$ in the case of the Rosetta molecular cloud, and there is no correlation between the atomic and molecular gas rotation axes. Their results demonstrate that giant molecular clouds do not simply inherit their present velocity field from the atomic gas from which they formed and suggest that a disordered or turbulent component may play an important role in setting the rotation axes. Our analysis of Taurus extends this finding to a smaller molecular cloud and shows that \sodium{} and \potassium{} resonance absorption lines can also be used to trace the structure of the atomic gas associated with molecular clouds. 
% If we interpret the difference in the \sodium{} and CO systemic velocity of $\sim 1$%\,km/s as due to turbulence we can use the turbulent structure function ($\delta v %%\sim 0.8 l^{0.5} $, \citealt{hb2004}) to measure the spatial displacement between the %atomic and molecular clouds in Taurus. 

%{\bf COMMENT from MH :If you want to include a physical mechanism, then magnetic %braking would produce misaligned/orthogonal rotational axes between cloud and %extended envelope}

\section{Conclusions and Perspectives}
Narrow \sodium{} absorption resonance lines have been long known to be common in the optical spectra of TTs. However, their origin remained elusive. Here, we have presented a detailed analysis of the \sodium{} and \potassium{} resonance lines toward nearly 40 TTs in Taurus spanning a range of evolutionary stages. Our main findings can be summarized as follows:

\begin{itemize}

\item The peak centroids of the \sodium{} at 5889.95\,\AA{} and \potassium{} at 7698.96\,\AA{}  are identical within our velocity uncertainty and their EWs are positively correlated. This demonstrates that the two transitions trace the same atomic gas.

\item Several of the \sodium{} and \potassium{} absorption lines have radial velocities substantially different from the stellar radial velocity and are detected even toward stars with no circumstellar disks. This demonstrates that \sodium{} and \potassium{} do not trace circumstellar disk gas, hence these features cannot be used to  investigate the Na/K ratio in disks, which is relevant for giant exoplanet atmospheres. 

\item The \sodium{} and \potassium{} radial velocities have a large spread ($\sim$10\,km/s) that cannot be accounted for by  gas in the local bubble.

\item The distribution of \sodium{} radial velocities in equatorial coordinates
shows a clear gradient suggesting that the absorption is associated with the Taurus molecular cloud. Assuming that the gradient is due to cloud rotation, the atomic gas rotates along an axis parallel to the magnetic field lines and prograde with respect to differential galactic rotation. This is different from the molecular gas which is known to have a retrograde motion.

\item  The atomic gas traced by \sodium{} and \potassium{} absorption lines is more extended and warmer than the molecular gas traced by low J CO rotational lines.

\end{itemize}

The almost anti-parallel rotation axis between the atomic and molecular gas cannot be explained by simple top-down formation scenarios according to which molecular clouds inherit the velocity field and angular momentum from the rotating galactic disk within which they form. A similar result is found by \citet{ib2011} when analyzing the rotational axes of the atomic and molecular gas from five giant molecular clouds in the Milky Way. In their study, they used the HI 21\,cm emission as a tracer of the atomic envelope of molecular clouds. Our results suggest that \sodium{} and \potassium{} absorption lines can also be used to trace that atomic gas. A direct comparison of these two different tracers would be extremely valuable. With the Gaia-ESO survey acquiring high-resolution UVES spectra of $\sim$5,000 stars in the Milky Way \citep{Smiljanic}, \sodium{} and \potassium{} absorption lines may be used to trace the kinematic of the atomic gas associated with the molecular gas of several other nearby star-forming regions.

\acknowledgments
The authors thank D. E. Welty for providing the high-resolution optical spectra toward HD~27778 and J. L. Pineda for sharing the Herschel spectra. I.P. would like to thank  T. Koskinen and P. Lavvas for stimulating discussions on giant exoplanet atmospheres and L. Hartmann and T. Megeath for helpful discussions on star formation processes. This work was partially supported by a NSF Astronomy \& Astrophysics Research Grant (ID: 1312962).


%% in parentheses, after the keyword, but they will not be verified.

{\it Facilities:} \facility{Keck}



\appendix

\section{Broad \sodium{} and \potassium{} absorption features}\label{sub:broadNa}
The five objects with broad \sodium{} absorptions are among the highest accretors in our sample ($\dot{M} \ge 10^{-8}$\,M$_\sun$/yr, e.g. Gullbring et al.~1998, White \& Ghez~2001, Hartmann et al.~1998).
As shown in Fig.~\ref{fig:naoi_jets}, broad absorptions are both on the blue and on the red side of the Na emission. A red absorption at several hundred km/s is a signature of mass infall (e.g. Edwards et al. 1994). Conversely, the broad absorptions located between -150 and -50\,km/s are most likely associated with material ejected from the star. Blueshifted absorptions are also present in the \potassium{} profiles of DG~Tau and DO~Tau but they are much weaker than those in the \sodium{} line. The inference of mass ejection is confirmed by the velocity coincidence of these absorption features with absorptions in the H$\alpha$ line and emission in the [OI] 6300\,\AA , a known tracer of jets/outflows (e.g. HEG95). We note that in the case of DR~Tau the absorption in the H$\alpha$ line is very shallow and difficult to see with the scaling used in Fig.~\ref{fig:naoi_jets} while the [OI] jet emission is more pronounced in the older and lower resolution spectra by HEG95, their Fig.~5.  A notable difference between the blue absorption in the \sodium{} and H$
\alpha$ lines is that the latter are broader, suggesting that Na traces only specific regions along the jet. In addition, the fact that these velocities are smaller than the peak velocities of the forbidden oxygen emission lines indicates that H$\alpha$ and \sodium{} absorptions mostly trace ambient material shocked to lower speeds.

\clearpage

\begin{figure}
\includegraphics[angle=0,scale=1.0]{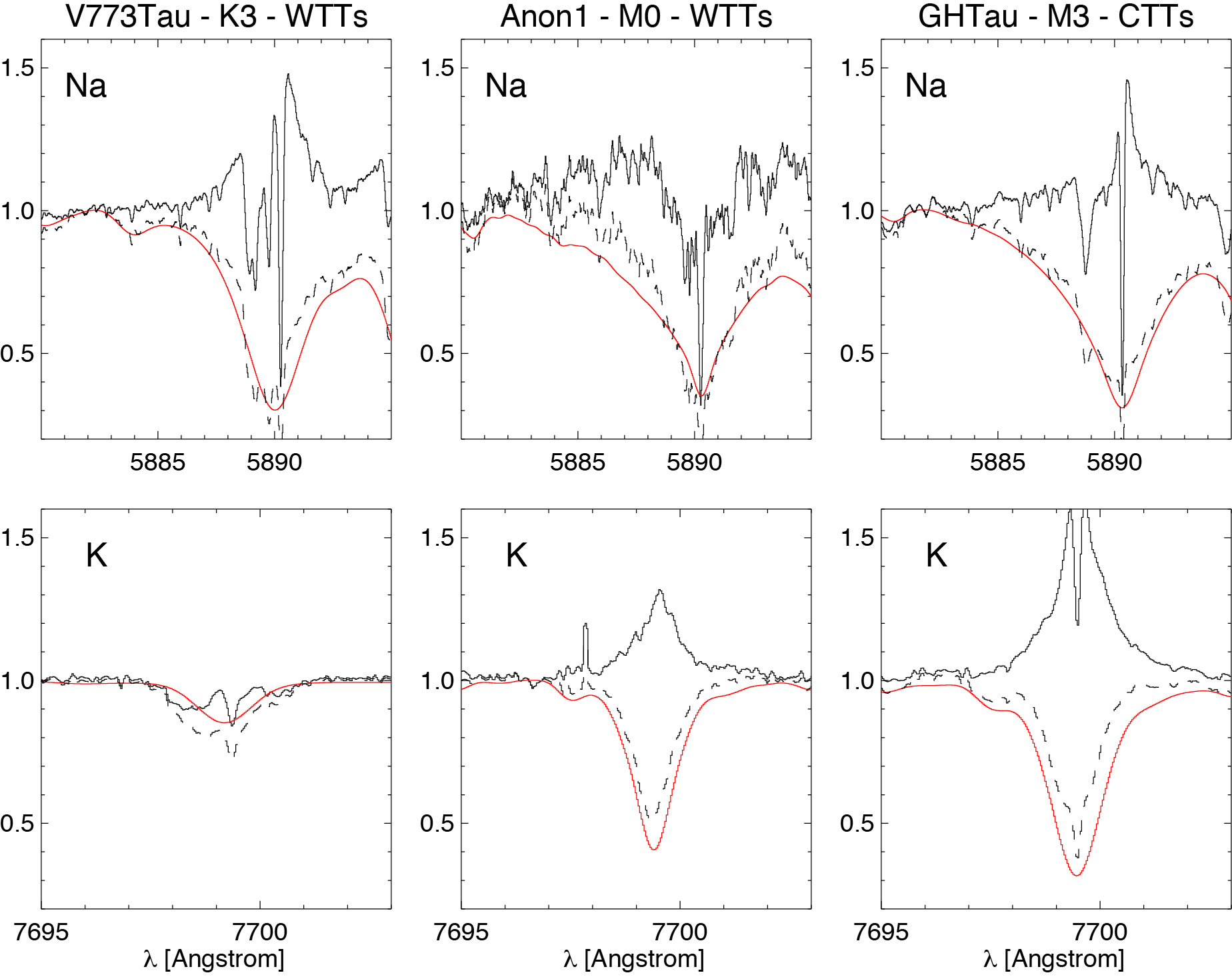}
\caption{Sample of original Na and K spectra (dashed line), scaled/veiled and broadened photospheric standard spectra (read solid line), and residual profiles (black solid line). These stars are chosen to cover the large range of SpTy of our sample and both accreting (CTTs) and non-accreting (WTTs) objects. Because Na and K lines are gravity sensitive, care should be taken in interpreting weak excess emission in the residual profiles, especially if the emission is as broad as that of photospheric lines. The figure also illustrates that the position and width of the narrow Na and K absorption features, which are the focus of this paper, are not affected by the photospheric subtraction. }
\label{fig:sample_resid}
\end{figure}

%%%%%%%%%%%%%%%%%%%%%%%%% Figure showing FWHMs
\begin{figure}
\includegraphics[angle=0,scale=1.0]{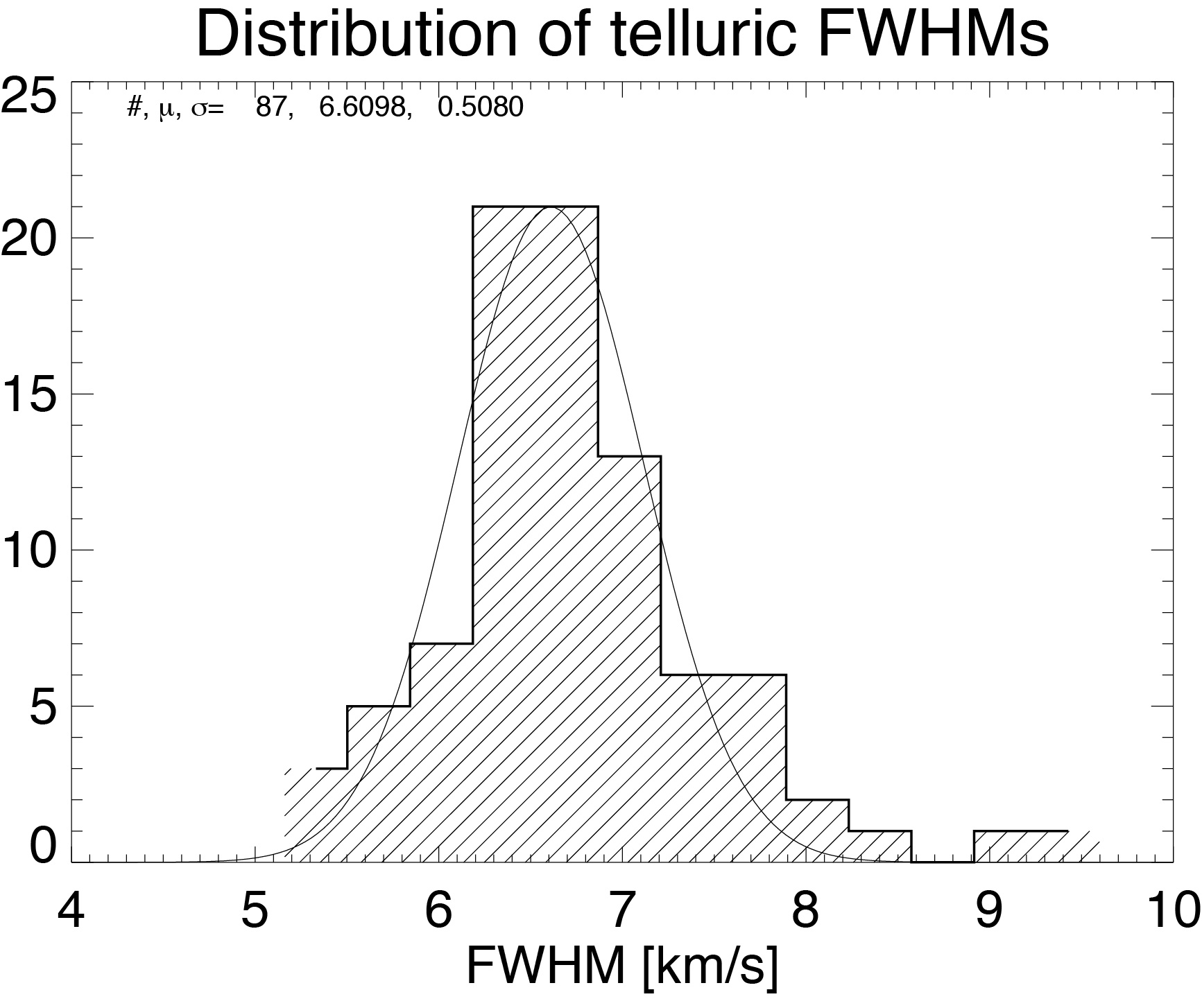}
\caption{Distribution of FWHMs from nearly 90 telluric lines. A gaussian fit to the distribution results in a mean FWHM of 6.6\,km/s, corresponding to a spectral resolution of $\sim$45,000 and standard deviation of 0.5\,km/s.}
\label{fig:fwhms}
\end{figure}

%%%%%%%%%%%% COMMENT: FZTAU IS NOT NEEDED BECAUSE NOW I SHOW ALL THE PROFILES
%\begin{figure}
%\includegraphics[angle=0,scale=1.0]{NaK_fig.ps}
%\caption{Residual normalized \sodium{} 5890 (black solid line), \sodium{} 5896 (blue dashed line), and %\potassium{} 7699\AA{} (red dot-dashed line) profiles for FZ~Tau. Note that the deep absorption at $\sim$%+20\,km/s is present in all three profiles, it is centered at the same velocity, and has the same width.}
%\label{fig:nakfig}
%\end{figure}
%%%%%%%%%%%% 

%%%%%%%%%%%%%%%%% START SUMMARY FIGURES OF PROFILES
\begin{figure}
\includegraphics[angle=0,scale=1.0]{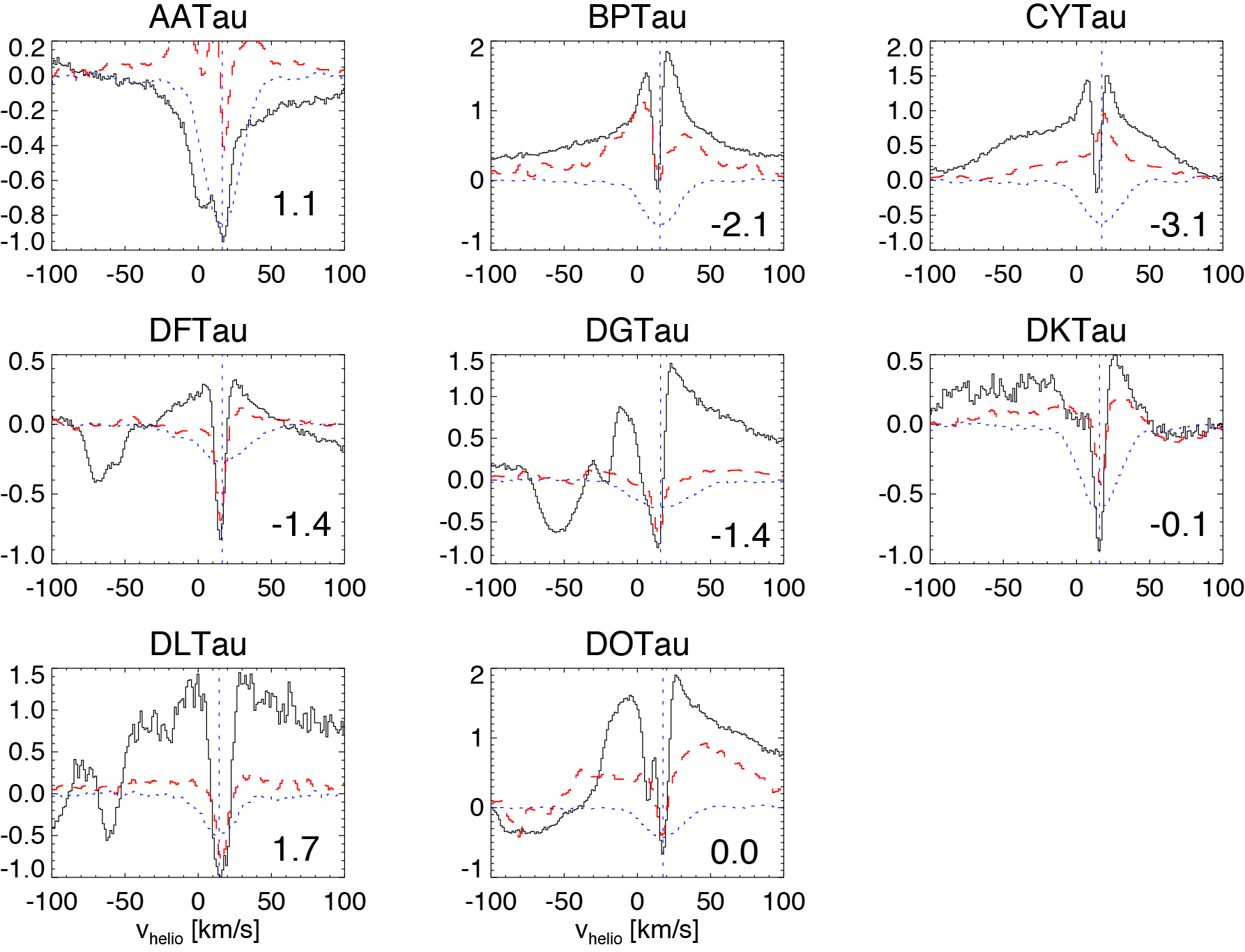}
\caption{Residual \sodium{} 5890 (black solid line) and \potassium{} 7699\AA{} (red dashed line) profiles for CTTs. The x-axis gives heliocentric velocities. 
Dotted (blue) lines show the Li photospheric profile at 6707.83\,\AA{} and its centroid. The number on the lower right corner of each panel gives the difference between the narrow \sodium{} absorption and the 
stellar radial velocity measured from the Li line. The \potassium{} profiles of BP~Tau and DO~Tau are multiplied by a factor of 5 while all others by a factor of 1.5.}
\label{fig:NaKLi_ctts1}
\end{figure}

\begin{figure}
\includegraphics[angle=0,scale=1.0]{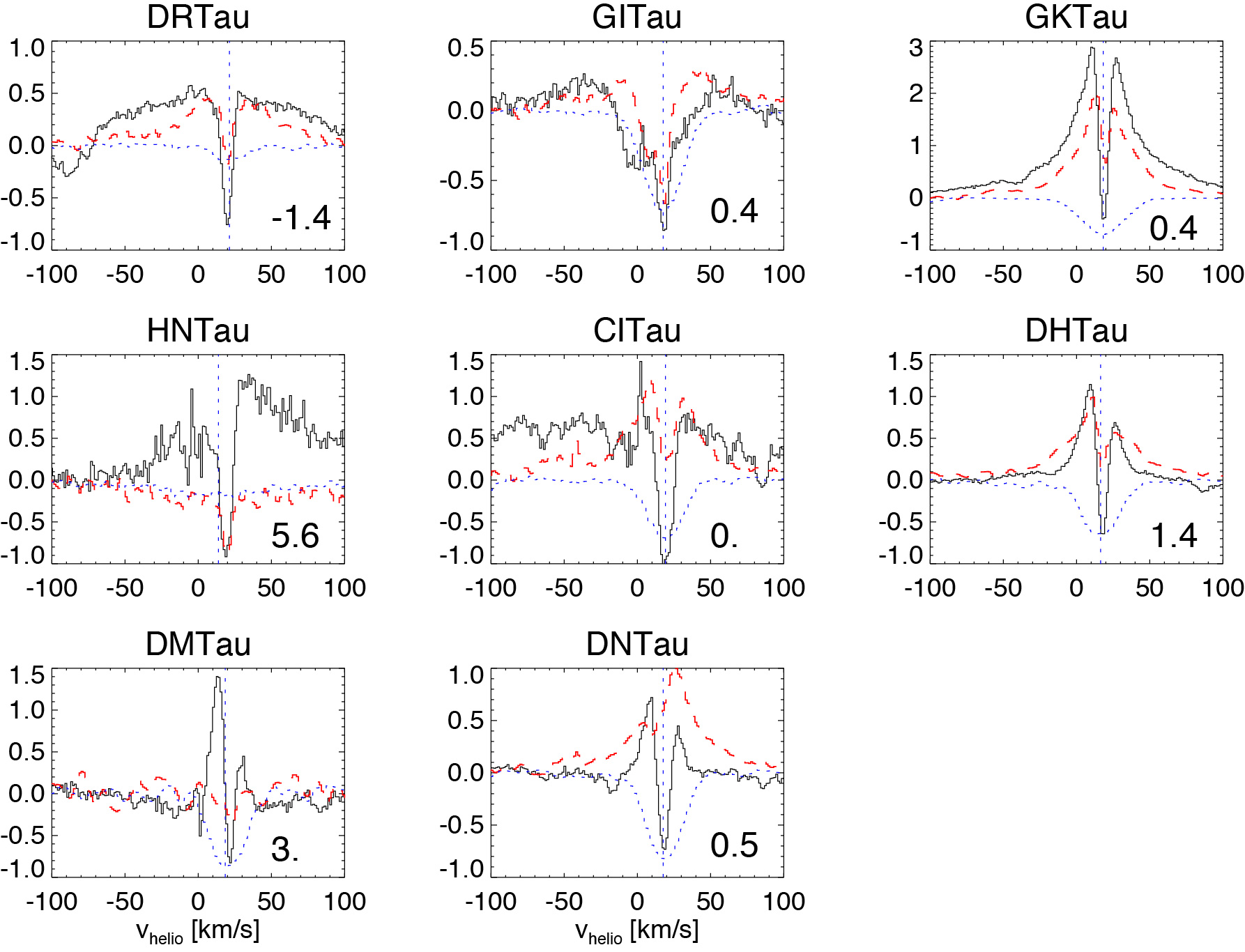}
\caption{Same as Fig.~\ref{fig:NaKLi_ctts1}. All \potassium{} profiles are multiplied by a factor of 2.}
\label{fig:NaKLi_ctts2}
\end{figure}

\begin{figure}
\includegraphics[angle=0,scale=1.0]{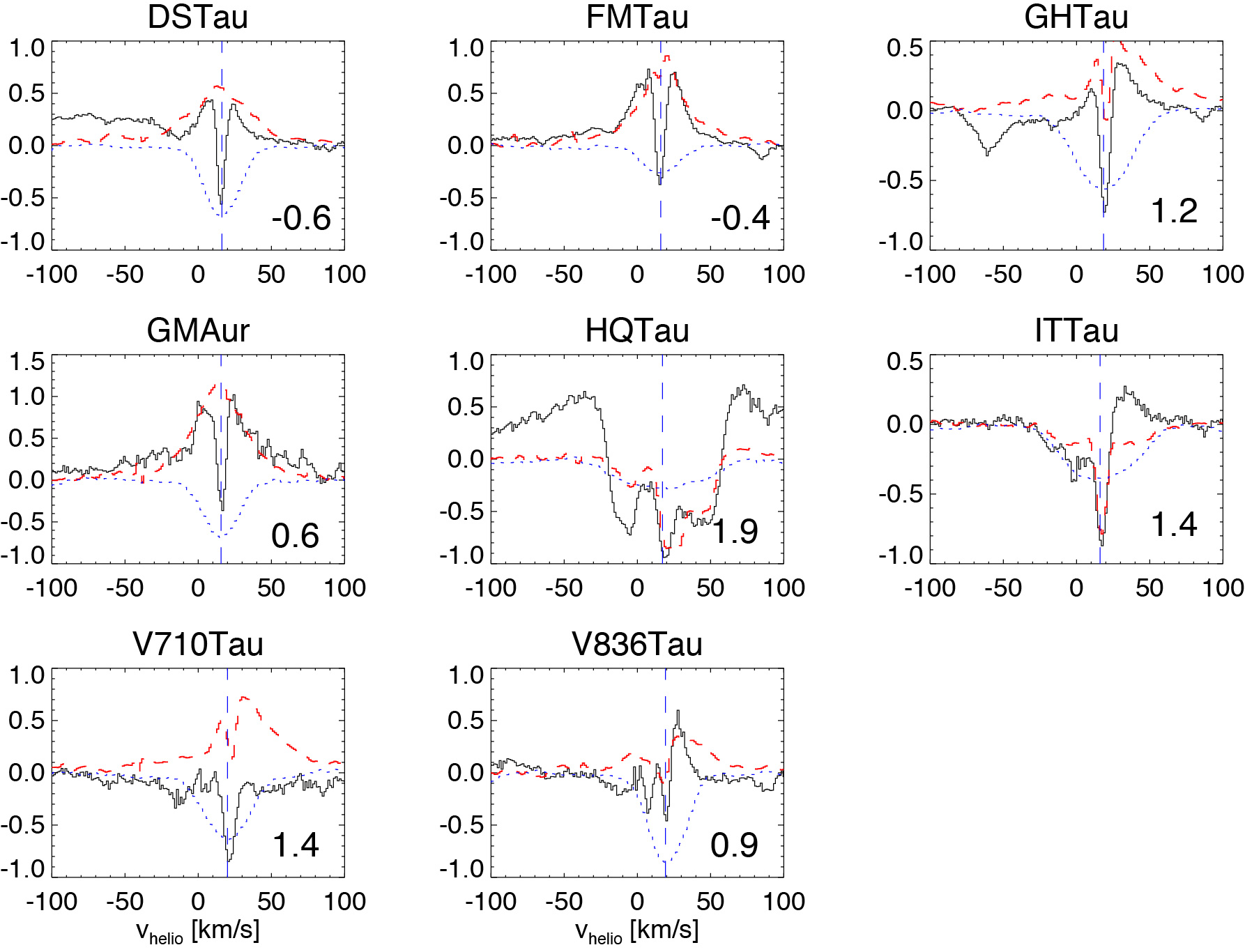}
\caption{Same as Fig.~\ref{fig:NaKLi_ctts1}. All \potassium{} profiles are multiplied by a factor of 1.5.}
\label{fig:NaKLi_ctts3}
\end{figure}

\begin{figure}
\includegraphics[angle=0,scale=1.0]{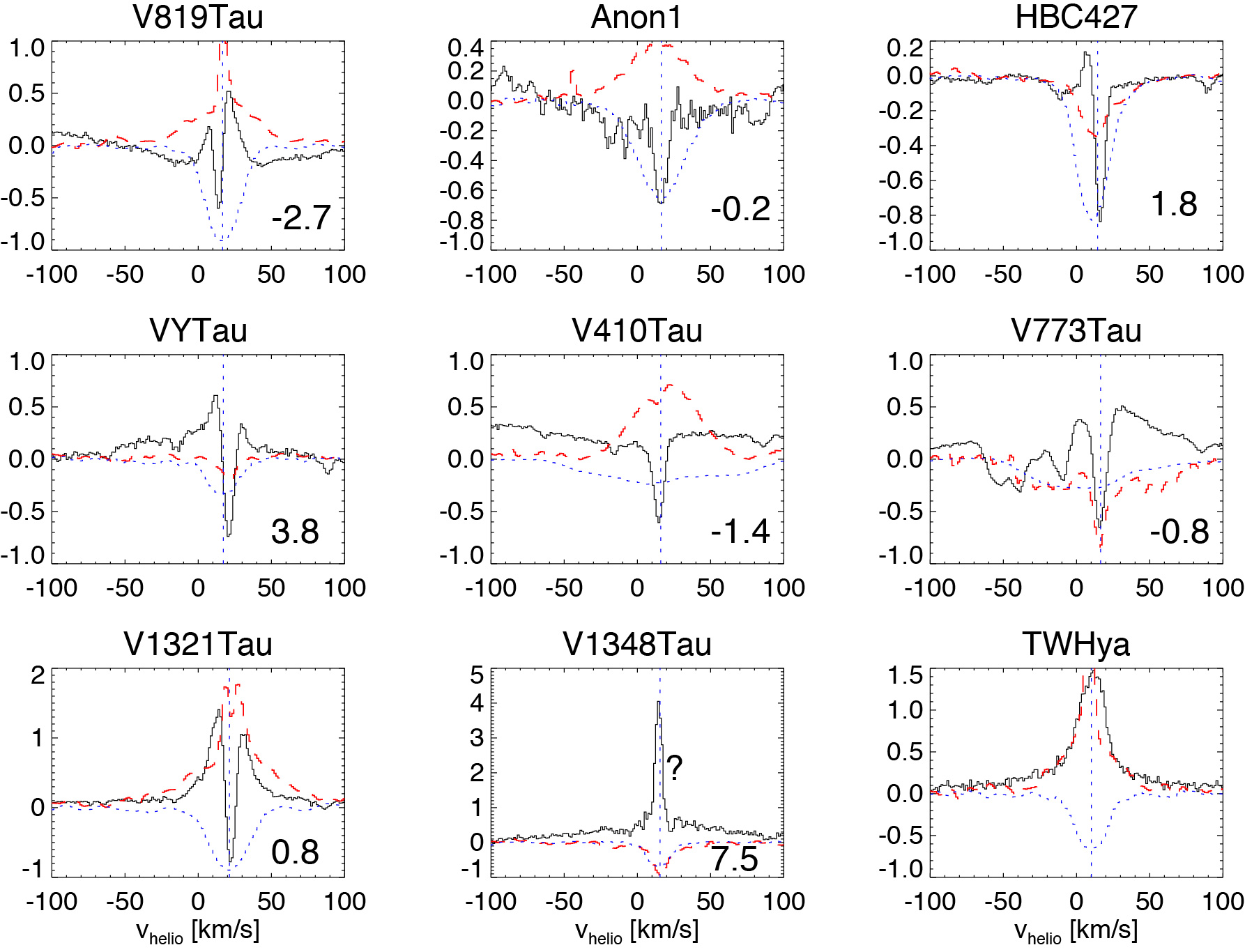}
\caption{Same as Fig.~\ref{fig:NaKLi_ctts1} but for the sample of WTTs and TW~Hya. The \potassium{} profiles of V410~Tau and V1321~Tau are multiplied by a factor of 2 while that of V773~Tau by a factor of 5.}
\label{fig:NaKLi_wtts}
\end{figure}

\begin{figure}
\includegraphics[angle=0,scale=1.0]{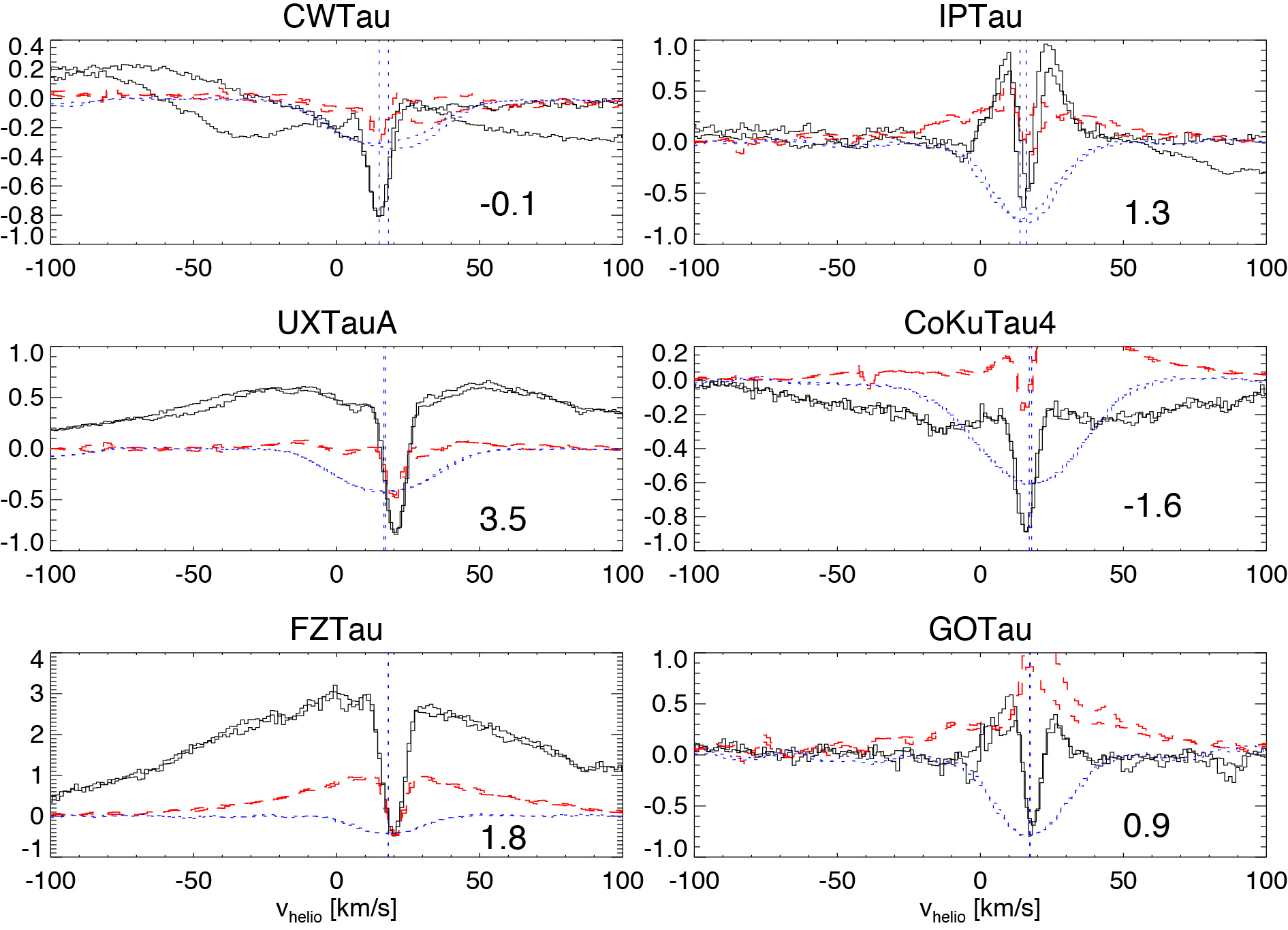}
\caption{Same as Fig.~\ref{fig:NaKLi_ctts1} but for sources that have multiple observations. The \potassium{} profiles of CoKu~Tau4 are multiplied by a factor of 2. The sharp absorptions in the \sodium{} and \potassium{}  are not time variable. Note that the IP~Tau 2006 \sodium{} and the Li heliocentric velocities are both $\sim$1.8\,km/s smaller than in the 2012 spectra, while the difference between the \sodium{} and Li lines remains the same in the two epochs. This shows that the absolute wavelength calibration can be off by up to $\sim$2\,km/s, perhaps depending on the source centering, as found in the comparison between our and literature radial velocities (Sect.~\ref{sect:analysis}).}
\label{fig:NaKLi_multi}
\end{figure}

%%%%%%%%%%%%%%%%% FIGURE HQTau-CITau
\begin{figure}
\includegraphics[angle=0,scale=1.0]{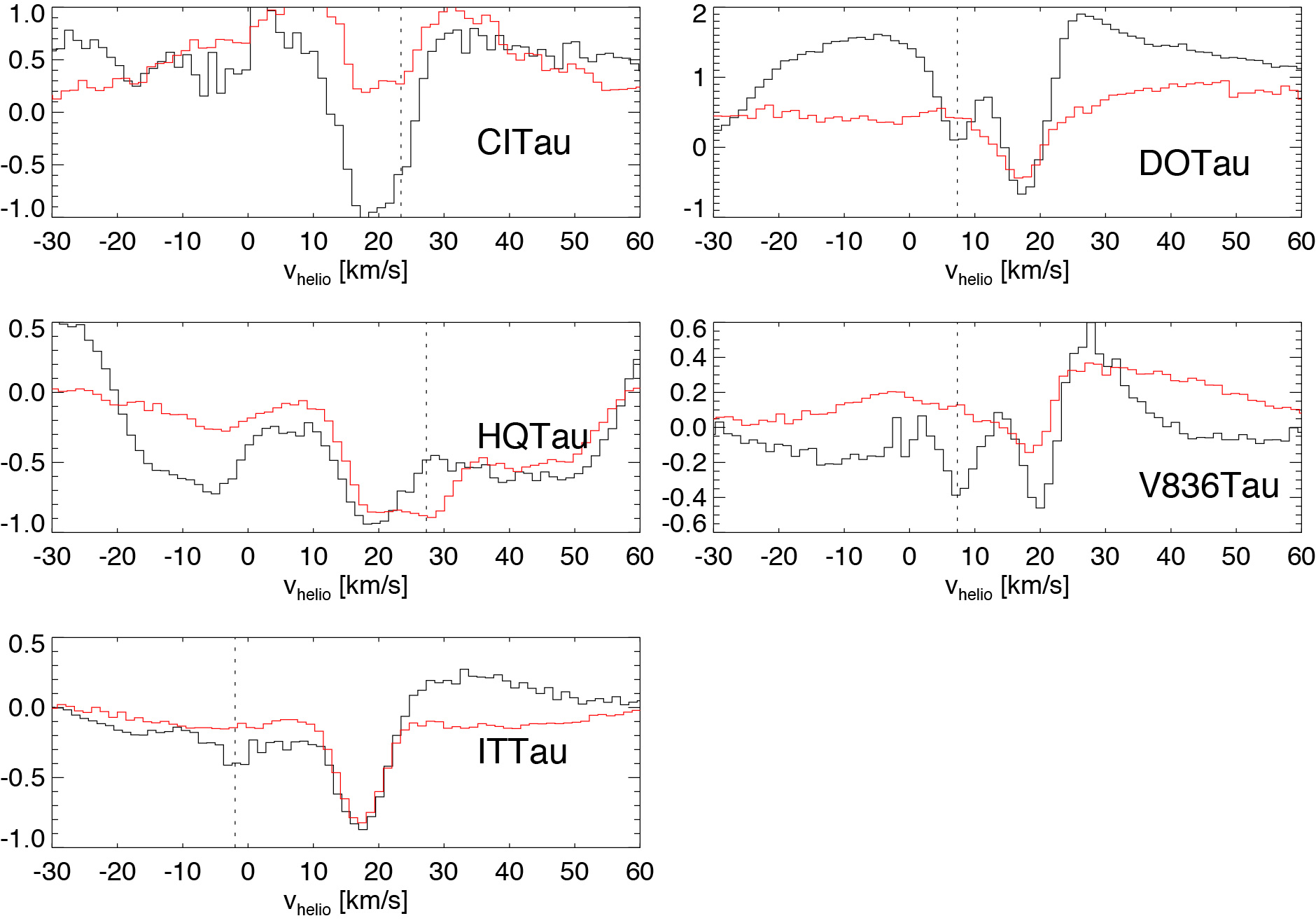}
\caption{Comparison of \sodium{} (black) and \potassium{} (red) residual profiles for the five TTs showing one additional narrow absorption feature in their spectra. As in previous figures the \potassium{} profiles are multiplied by a factor of 5 for DO~Tau, a factor of 2 for CI~Tau, and a factor of 1.5 for the other three sources. The location of the additional narrow component is indicated with a dotted line. We do not include these additional narrow components in any of our analysis.
}
\label{fig:hqtau-citau}
\end{figure}
%%%%%%%%%%%%%%%%% 

%%%%%%%%%%%%%%%%% FIGURE NaK comparison
\begin{figure}
\includegraphics[angle=0,scale=1.0]{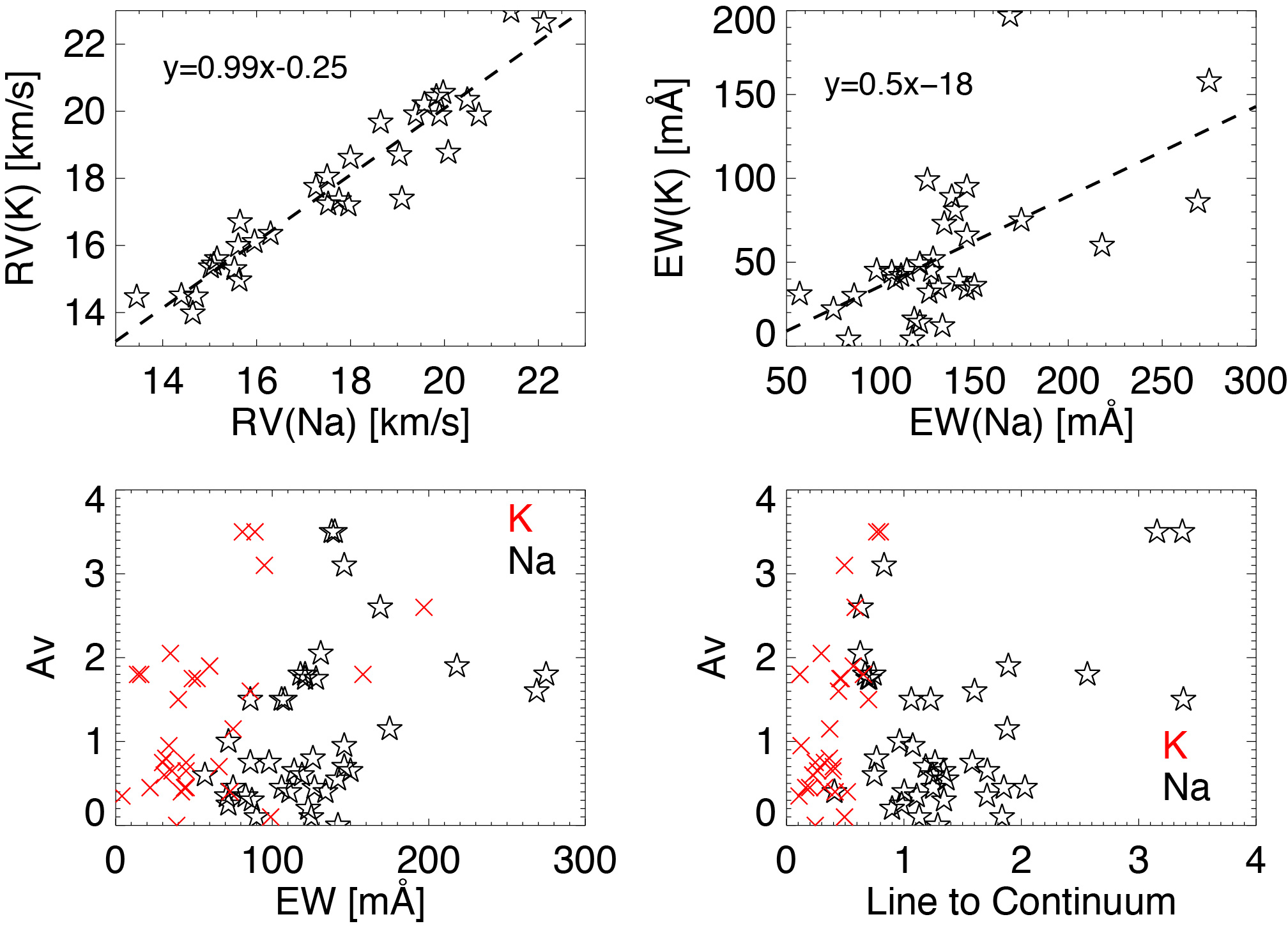}
\caption{Upper panels: Comparison of \sodium{} and \potassium{} peak centroids (left) and EWs (right). We only plot sources where both narrow absorption lines are detected. Both quantities are positively correlated. In both panels a dashed line shows the best linear fit.
Lower panels: Source extinction A$_{\rm V}$ as a function of EW (left) and line-to-continuum ratio (right). Detected \sodium{} absorptions are shown with stars while \potassium{} absorptions with an X. The EWs of \sodium{} lines are about 2 times larger than those of \potassium{} lines.
}
\label{fig:NaK_comp}
\end{figure}
%%%%%%%%%%%%%%%%% 

%%%%%%%%%%%%%%%%% FIGURE HISTOGRAM OF NA, CO AND LI PEKS AND WIDTHS
%\begin{figure}
%\includegraphics[angle=0,scale=1.0]{Histo_Li_Na_CO.ps}
%\caption{Distribution of radial velocities (upper panel) and line FWHMs (lower panel).
%Histograms filled with diagonal lines are for the Na absorption (sample overlapping with the CO %maps) while (black) dashed lines are for the all Keck sample. Dash-dot lines (red) are for the CO %emission profiles. Note that the radial velocities of the CO lines cover a smaller range than %those of the \sodium{} lines.
%}
%\label{fig:Histo_Li_Na_CO}
%\end{figure}

%NEW SUCH FIGURE
\begin{figure}
\includegraphics[angle=0,scale=1.0]{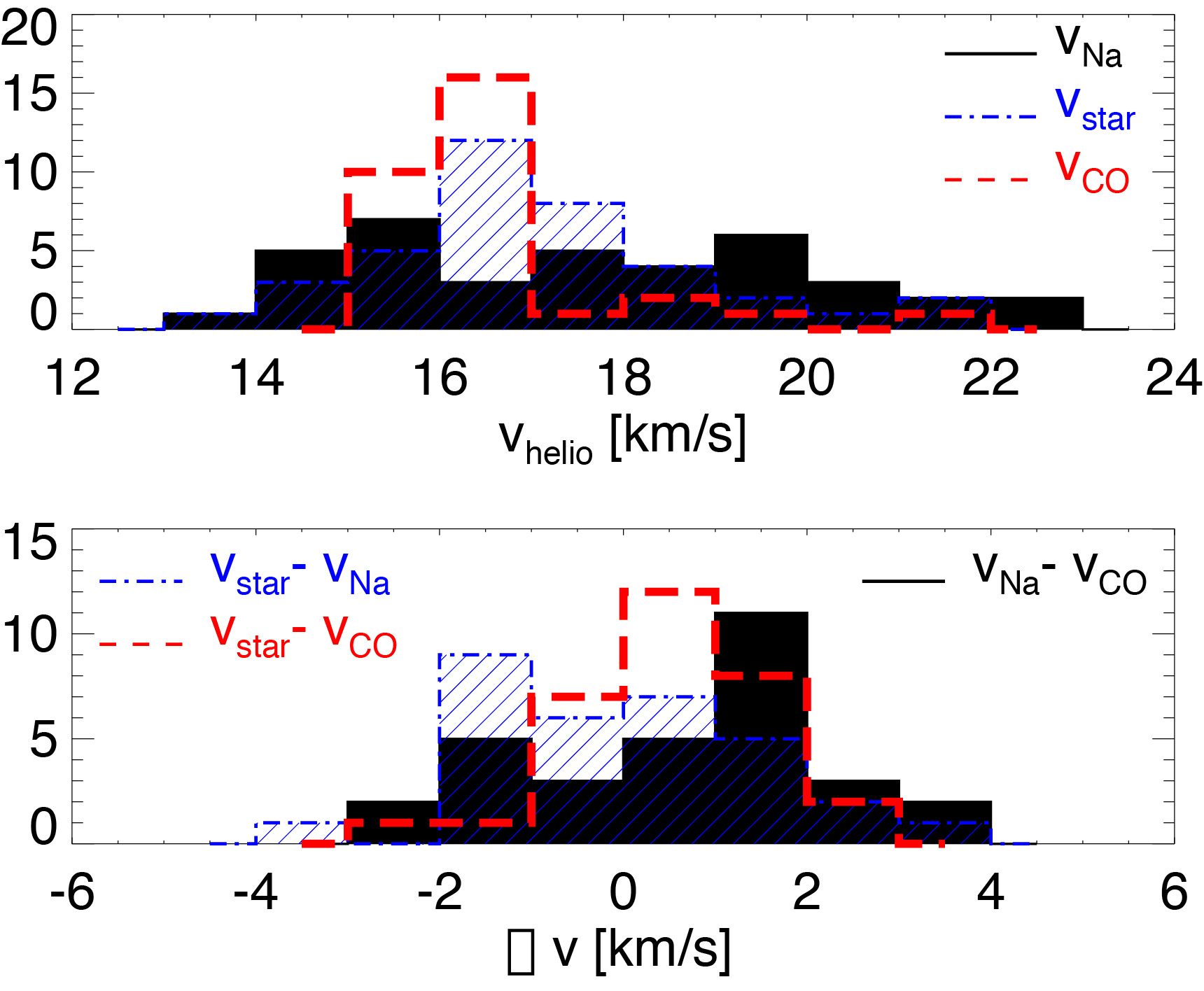}
\caption{Distribution of heliocentric radial velocities (upper panels) and velocity differences (lower panel). For the molecular gas we only include sources for which the 
signal-to-noise of the $^{12}$CO spectra is larger than 5 (see last column of Tables~\ref{tab:RV2006} and \ref{tab:RV2012}).}
\label{fig:Histo_Li_Na_CO}
\end{figure}

%%%%%%%%%%%%%%%%% FIGURE GROUPINGS
%\begin{figure}
%\includegraphics[angle=0,scale=1.0]{groups.ps}
%\caption{Figure showing the Na~D absorption profiles of sources that are closer than 0.5$\deg%$ to each other. In general, we see a good match between line centroids and full widths for %sources in the same group. Note that the Na absorption toward DG~Tau is clearly affected by %the jet powered by this star, see text.
%}
%\label{fig:groups}
%\end{figure}
%%%%%%%%%%%%%%%%% 

%%%%%%%%%%%%%%%%% FIGURE LOCATION OF SOURCES IN VELOCITIES
\begin{figure}
\includegraphics[angle=0,scale=1.0]{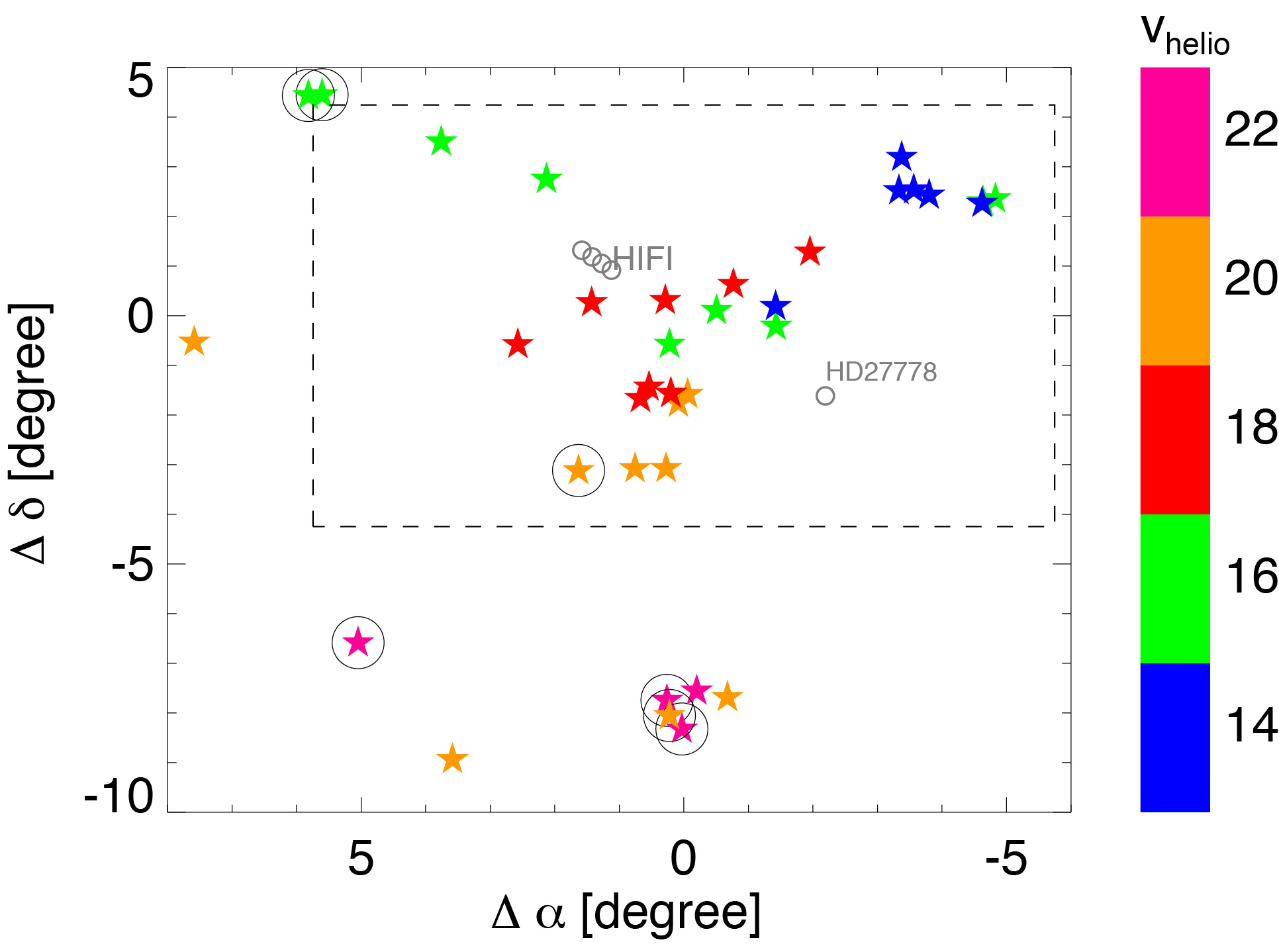}
\caption{Distribution of our TTs (stars) in relative equatorial coordinates. The center of the plot is the center of the CO map \citep{narayanan08}, its extent is shown by the dashed line rectangle.
TTs are color coded based on their \sodium{} heliocentric RVs: each color bin covers 2\,km/s and the mean RV for each color bin is listed next to the colorbar. We also plot the location of the background B3 star HD~27778 studied by \cite{wh2001} and of the Herschel/HIFI pointings by \citet{orr2014}. The signal-to-noise of the $^{12}$CO CfA spectra toward circled stars is less than 5. As mentioned in the text and Fig.~\ref{fig:NaKLi_wtts} the detection of \sodium{} absorption toward V1348~Tau is not secure. Note that the \sodium{} RVs increase going from NW to SE. }
\label{fig:colormap}
\end{figure}
%%%%%%%%%%%%%%%%% 

%%%%%%%%%%%%%%%%% FIGURE POSITION-VELOCITY MAP
\begin{figure}
\includegraphics[angle=0,scale=1.0]{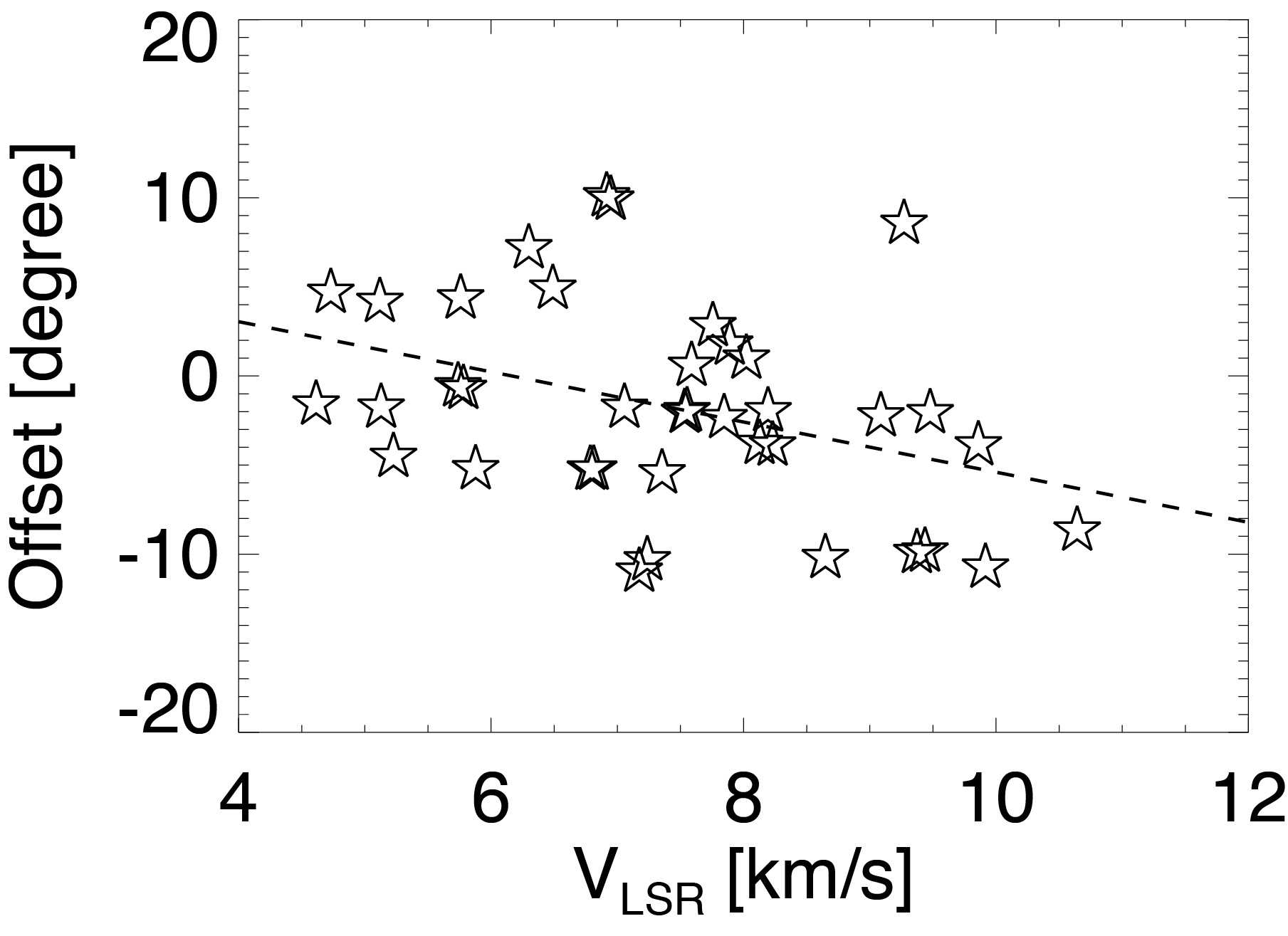}
\caption{Perpendicular offset of each TTs from the Na cloud rotation axis (P.A.=210$^\circ$) as a function of the \sodium{} LSR velocity. The linear trend between the two quantities shows that a plane is a good fit to the velocities. }
\label{fig:position_velocity}
\end{figure}
%%%%%%%%%%%%%%%%% 

%%%%%%%%%%%%%%%%% FIGURE LOCATION OF SOURCES IN VELOCITIES on top of CO map
\begin{figure}
\includegraphics[angle=0,scale=1.0]{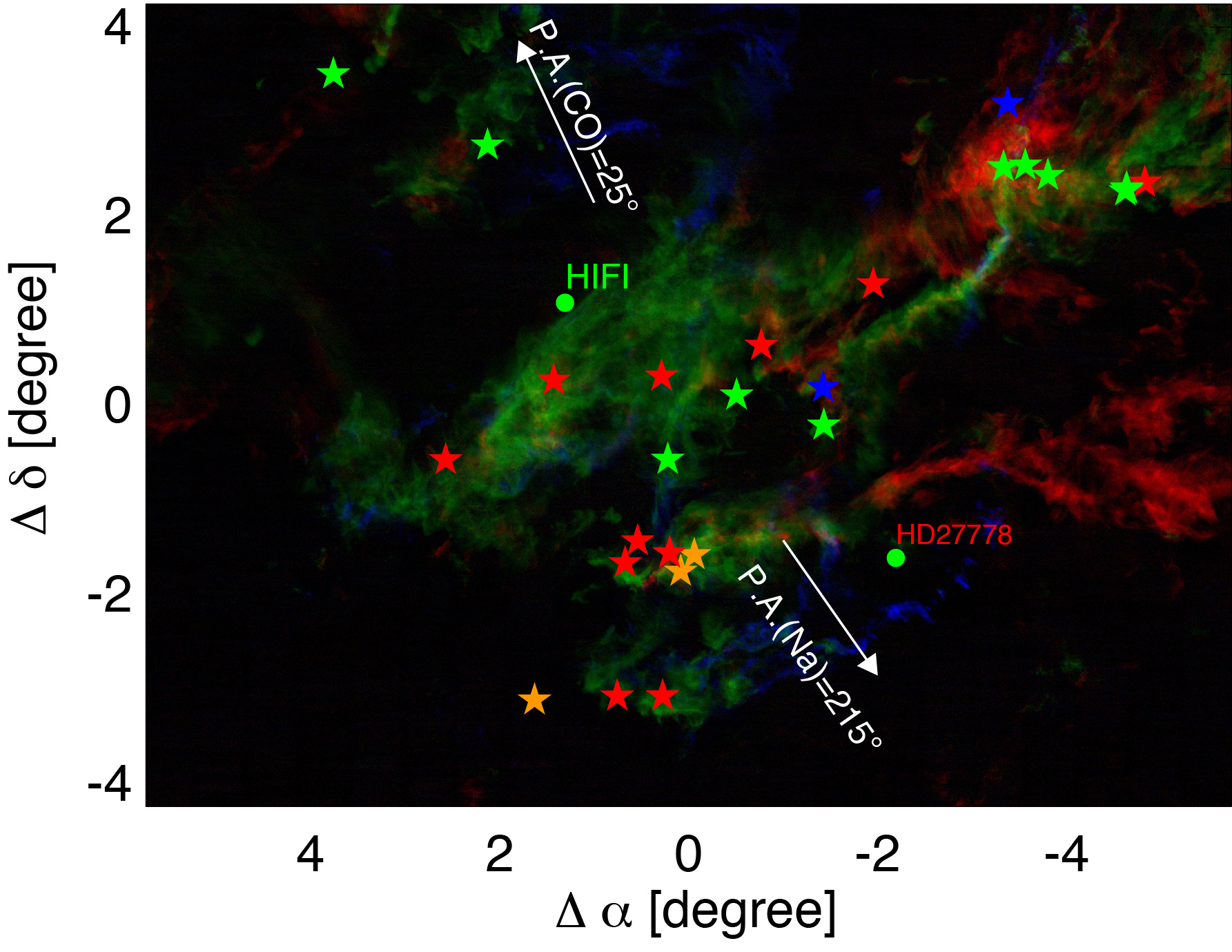}
\caption{Color-coded image of the $^{13}$CO integrated intensity in three LSR velocity intervals: 3-5\,km/s is coded blue, 5-7\,km/s green, and 7-9\,km/s red, see also 
\citet{goldsmith2008}. Our TTs are shown with stars, the color coding representing their \sodium{} RV also in LSR. We included one additional velocity bin (9-11\,km/s LSR, orange) to cover the larger velocity range seen in the \sodium{} and \potassium{} absorption lines. White arrows show the position angles of $^{13}$CO and the \sodium{} rotation axis, velocity gradients are perpendicular to each axis. Note that the $^{13}$CO and the \sodium{} axes are almost anti-parallel. The magnetic field direction is at position angle 25$^\circ$ (205$^\circ$), very similar to the rotation axis of the molecular and of the \sodium{} gas.}
\label{fig:colormap13CO}
\end{figure}
%%%%%%%%%%%%%%%%% 

%%%%%%%%%%%%%%%%% FIGURES FOR NA BROAD ABS
\begin{figure}
\includegraphics[angle=0,scale=1.0]{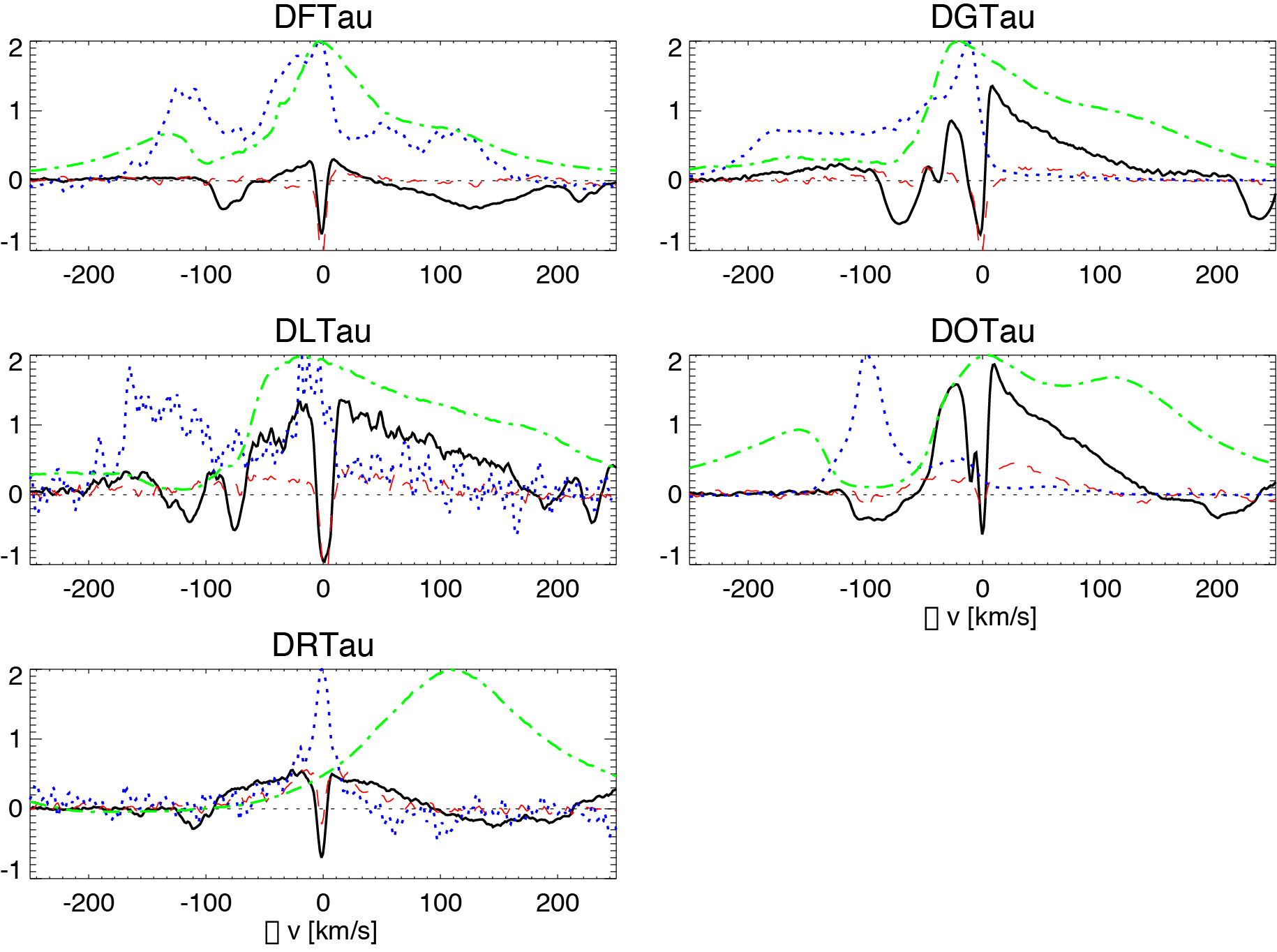}
\caption{Sources with broad \sodium{} absorptions. 
A black solid line shows the residual \sodium{} 5890\,\AA{} profile while a red dashed line the \potassium{} line at 7699\AA.
Normalized residual profiles for the H$\alpha$ (green dash-dot line) and the [\oi ] 6300\,\AA{} 
(blue dotted line) lines are also superimposed.
The velocity scale is larger than in previous figures to show both the broad blue- and red-shifted absorptions. All velocities are in the stellocentric reference frame.
}
\label{fig:naoi_jets}
\end{figure}
%%%%%%%%%%%%%%%%% 

%% If you are not including electonic art with your submission, you may
%% mark up your captions using the \figcaption command. See the
%% No more than seven \figcaption commands are allowed per page,
%% so if you have more than seven captions, insert a \clearpage

%% Tables should be submitted one per page, so put a \clearpage before

%% Two options are available to the author for producing tables:  the
%% deluxetable environment provided by the AASTeX package or the LaTeX
%% table environment.  Use of deluxetable is preferred.
%%

%% Three table samples follow, two marked up in the deluxetable environment,
%% one marked up as a LaTeX table.

%% In this first example, note that the \tabletypesize{}
%% command has been used to reduce the font size of the table.
%% We also use the \rotate command to rotate the table to
%% landscape orientation since it is very wide even at the
%% Note also that the \label command needs to be placed

%% This table also includes a table comment indicating that the full
%% version will be available in machine-readable format in the electronic

\clearpage

\begin{deluxetable}{ccc}
\tabletypesize{\scriptsize}
%\rotate
\tablecaption{2006 campaign: observation log\label{tab:obs2006}}
\tablewidth{0pt}
\tablehead{
\colhead{Source} & \colhead{2MASSJ} & \colhead{Exposure Time}  
}
\startdata
AA~Tau & 04345542+2428531 & 600  \\

BP~Tau & 04191583+2906269 & 600  \\

CW~Tau & 04141700+2810578 & 2700  \\

CY~Tau & 04173372+2820468 & 900 \\

DF~Tau & 04270280+2542223 & 300 \\

DG~Tau & 04270469+2606163 & 300  \\
          
DK~TauA & 04304425+2601244 & 300  \\

DL~Tau & 04333906+2520382 & 600 \\

DO~Tau & 04382858+2610494 & 1200  \\

DR~Tau & 04470620+1658428 & 900  \\

GI~Tau & 04333405+2421170 & 900  \\

GK~TauA & 04333456+2421058 & 900  \\

HN~TauA & 04333935+1751523 & 900  \\

IP~Tau & 04245708+2711565 & 300  \\

TW~Hya & 11015191-3442170 & 600  \\

UX~TauA & 04300399+1813493 & 1200  \\

V819~Tau & 04192625+2826142 & 900  \\
\enddata
\end{deluxetable}

\begin{deluxetable}{ccc}
\tabletypesize{\scriptsize}
%\rotate
\tablecaption{2012 campaign: observation log\label{tab:obs2012}}
\tablewidth{0pt}
\tablehead{
\colhead{Source} & \colhead{2MASSJ} & \colhead{Exposure Time}  
}
\startdata
Anon~1 & 04132722+2816247  & 1480 \\

CI~Tau & 04335200+2250301 & 2121  \\

CoKu~Tau4 & 04411681+2840000 & 7200 \\

DH~TauA & 04294155+2632582 & 3000 \\

DM~Tau & 04334871+1810099 & 3600 \\

DN~Tau & 04352737+2414589 & 899  \\

DS~TauA & 04474821+2925138 & 900  \\
          
FM~Tau & 04141358+2812492 & 3600 \\

FZ~Tau & 04323176+2420029 & 7200  \\

GH~Tau & 04330622+2409339 & 1500   \\

GM~Aur & 04551098+3021595 & 900  \\

GO~Tau & 04430309+2520187 & 6202   \\

HBC~427 & 04560201+3021037 & 600  \\

HQ~Tau & 04354733+2250216 & 900  \\

IP~Tau & 04245708+2711565 & 1500   \\

IT~TauA & 04335470+2613275 & 3710  \\

%RXJ0405 & 04051233+2632438 &  \\

UX~TauA & 04300399+1813493 & 300   \\

V1321~Tau & 04325323+1735337 & 2603 \\

V1348~Tau & 04525707+1919504& 600 \\

V410~Tau & 04183110+2827162 &  300\\

VY~Tau  & 04391741+2247533 & 300  \\

V710~TauA & 04315799+1821350 & 3600   \\

V773~Tau & 04141291+2812124 & 300  \\

V836~Tau & 05030659+2523197 & 2400  \\
\enddata
\end{deluxetable}

\begin{deluxetable}{lcccccc}
\tabletypesize{\scriptsize}
%\rotate
\tablecaption{2006 campaign: source properties relevant to this study\label{tab:psou2006}}
\tablewidth{0pt}
\tablehead{
\colhead{Source} & \colhead{SpTy}& \colhead{Multiplicity\tablenotemark{a}} & \colhead{REF} & \colhead{TTs} & \colhead{SED\tablenotemark{b}} & \colhead{REF} 

}
\startdata
AA~Tau & M0.6 & s & 1,2 & C & II & 2,3\\

BP~Tau & M0.5 &  s & 1,2 & C & II & 2,3\\

CW~Tau & K3 & s & 1,2 & C & II & 2,3 \\

CY~Tau & M2.3 & s & 1,2 & C & II & 2,3 \\

DF~Tau & M2.7 & 0.09 & 1,2 & C & II & 2,3 \\

DG~Tau & K7.0 & s & 1,2 & C & II & 2,3 \\
          
DK~TauA & K8.5 & 2.30 & 1,2 & C & II & 2,3 \\

DL~Tau & K5.5 & s & 1,2 & C & II & 2,3   \\

DO~Tau & M0.3 & s & 1,2 & C & II & 2,3 \\

DR~Tau & K6 & s & 1,2 & C & II & 2,3 \\

GI~Tau & M0.4 & s & 1,2 & C & II & 2,3 \\

GK~Tau & K6.5 & s & 1,2 & C & II & 2,3 \\

HN~TauA & K3 & 3.11 & 1,2 & C & II & 2,3 \\

IP~Tau & M0.6 & s & 1,2 & C & II & 2,3  \\

UX~TauA & K0.0 &  2.63,5.86 & 1,2 & C & T & 2,4  \\

V819~Tau & K8.0 & s & 1,2 & W & II/III\tablenotemark{c} & 2,3\\

\hline

TW~Hya &  M0.5 & s  & 1,5 & C & T & 6,7\\
\enddata
%% Text for table notes should follow after the \enddata but before
%% the \end{deluxetable}. Make sure there is at least one \tablenotemark
%% in the table for each \tablenotetext.
%\tablecomments{ }
\tablerefs{(1) Herczeg \& Hillenbrand~2014; (2) White \& Ghez~2001; (3) Kenyon \& Hartman 1995; (4) Furlan et al. 2009a;
(5) Rucinski et al.~2008; (6) Dupree et al.~2012; (7) Calvet et al.~2002 }
\tablenotetext{a}{For multiple systems we give the separation in arcseconds from the primary.}
\tablenotetext{b}{Known transition disks are marked with ’T’.}
\tablenotetext{c}{This source has a weak infrared excess at wavelengths longer than 10\,\micron , it may be transitioning from Class II to III but the excess is not as large as for classical transition disks (Furlan et al. 2009b).}
\end{deluxetable}

\begin{deluxetable}{lcccccc}
\tabletypesize{\scriptsize}
%\rotate
\tablecaption{2012 campaign: source properties relevant to this study\label{tab:psou2012}}
\tablewidth{0pt}
\tablehead{
\colhead{Source}& \colhead{SpTy} & \colhead{Multiplicity\tablenotemark{a}} & \colhead{REF} & \colhead{TTs} & \colhead{SED\tablenotemark{b}} & \colhead{REF} 
}
\startdata
Anon~1 &  M0.5 & s & 1,2 & W & III & 2,3  \\

CI~Tau & K5.5 & s\tablenotemark{c} & 1,2 & C & II & 2,3 \\

CoKu~Tau4 & M1.1 & 0.05 & 1,4 & W & T & 5,6\\

DH~Tau & M2.3 & s & 1,2 & C & II & 2,3  \\

DM~Tau & M3.0 & s & 1,2 & C & T & 2,6 \\

DN~Tau & M0.3 & s & 1,2 & C & II & 2,3 \\

DS~Tau & M0.4 & s & 1,2 & C & II & 2,3  \\
          
FM~Tau & M4.5 & s & 1,2 & C & II & 2,3 \\

FZ~Tau & M0.5 & s & 1,2 & C & II & 2,3 \\

GH~Tau & M2.3 & 0.31 & 1,2 & C & II & 2,3  \\

GM~Aur & K6.0 & s & 1,2 & C & T & 2,6 \\

GO~Tau & M2.3 & s & 1,2 & C & II & 2,3 \\

HBC~427 & K6.0 & 0.03 & 1,7 & W & II/III\tablenotemark{d}& 3,8 \\

HQ~Tau & K2.0 & s & 1,9  & C & II & 10,11\\

IP~Tau & M0.6 & s & 1,2 & C & II & 2,3 \\

IT~TauA & K6.0 & 2.39 & 1,2 & C & II & 2,3\\

%NOTE on RXJ0405.1+2632:
%I removed it because it is off CO map, we don’t know anything about this star
% and the disk and I don’t manage to clean the photosphere
%RXJ0405.1+2632 &  s & 5a & not in Kraus \\ 
%% HD 283323 this is RXJ0405

UX~TauA & K0.0 & 2.63,5.86 &  1,2 & C & T & 2,6 \\

VY~Tau  & M1.5 & 0.66 & 1,2 & W & II & 2,12 \\

V410~Tau &  K7 & 0.07 &  10,2 & W & III\tablenotemark{e} & 2,12\\

V710~TauA & M3.3 & 3.17 & 1,2  & C & II & 2,3 \\

V773~Tau & K4.0 &  0.1\tablenotemark{f} &  1,13  & W & II & 2,11 \\

V836~Tau &  M0.8 & s & 1,9 & C & II & 3,14,11\\

%V1321~Tau(RX J0432.8+1735)
V1321~Tau & M2 &  s & 10 &  W & III\tablenotemark{g} & 15,16\\
   
%V1348~Tau(RX J0452.9+1920) & s & 9 & W &  III & 9,16 \\
V1348~Tau & K5 & s & 10 & W &  III & 10,17 \\
\enddata
%% Text for table notes should follow after the \enddata but before
%% the \end{deluxetable}. Make sure there is at least one \tablenotemark
%% in the table for each \tablenotetext.
%%\tablecomments{Sources that have been observed more than once in our observational campaign have multiple velocities.}
\tablerefs{(1) Herczeg \& Hillenbrand~2014; (2) White \& Ghez 2001; (3) Kenyon \& Hartmann 1995; (4) Ireland \& Kraus 2008; (5) Cohen \& Kuhi 1979; (6) Furlan et al. 2009a; (7) Steffen et al. 2001; (8) Kenyon et al. 1998; (9) Kraus et al. 2012 ; (10) Nguyen et al. 2009; (11) Furlan et al. 2011; (12) Furlan et al. (2006); (13) Ghez et al. 1993; (14) White \& Hillebrand 2004; (15) Martin\& Magazzu 1999; (16) Cieza et al. 2013; 
(17) Wichmann et al. 1996}
\tablenotetext{a}{For multiple systems we give the separation in arcseconds from the primary and the flux ratio to the primary when known.}
\tablenotetext{b}{Known transition disks are marked with ’T’.}
\tablenotetext{c}{Suspected SB1, Nguyen et al. (2012).}
\tablenotetext{d}{This source has a weak infrared excess at wavelengths longer than 10\,\micron , it may be transitioning from Class II to III but the excess is not as large as for classical transition disks (Furlan et al. 2006)}
\tablenotetext{e}{Esplin et al.~(2014) report a small excess in the WISE W2 band (4.6\,\micron ) but note that the colors of V410~Tau are consistent with the stellar photosphere at longer wavelengths as found in Furlan et al.~(2006)}
\tablenotetext{f}{The quoted separation is between the A and B components. A is a short period (51 days) double-lined spectroscopic binary (Welty 1995). In addition Duchene et al.~(2003) and Woitas~(2003) identified an additional infrared component (C) within 0.3\arcsec{} from A.}
\tablenotetext{g}{Cieza et al.~(2013) report excess at wavelengths longer than 24\,\micron{} for V1321~Tau (also known as RXJ0432.8+1735). This source is thought to have a warm debris disk.}
\end{deluxetable}

%%%%%%%%%% Tables reporting stellar radial velocities and FWHM for the Li line

\begin{deluxetable}{c c ccc ccc cc}
\tabletypesize{\scriptsize}
%\rotate
\tablecaption{2006 campaign: Heliocentric radial velocities, widths and EWs of selected lines.\label{tab:RV2006}}
\tablewidth{0pt}
\tablehead{
\colhead{} & \colhead{Stellar\tablenotemark{a}} & \multicolumn{3}{c}{Na} & \multicolumn{3}{c}{K}&\multicolumn{2}{c}{$^{12}$CO}\\
\colhead{Source} & \colhead{RV}  & \colhead{RV} & \colhead{FWHM} &\colhead{EW} & \colhead{RV} & \colhead{FWHM} & \colhead{EW} &\colhead{RV} & \colhead{S/N}\\
\colhead{ } & \colhead{[km/s]} & \colhead{[km/s]} & \colhead{[km/s]} & \colhead{[m\AA]}&\colhead{[km/s]} & \colhead{[km/s]}& \colhead{[m\AA]} &\colhead{[km/s]} & \colhead{}
}
\startdata
AA~Tau &  16.4    & 17.5  &      8.8 & 134 & 18.1&5.6& 73 & 16.2 & 28\\

BP~Tau &  15.5   & 13.4 &      6.0 & 75 & 14.5&5.7& 22 & 16.1 & 15\\

CW~Tau & 18.1,14.8\tablenotemark{b}  & 15.1,14.7 &   8.0,7.7 & 121,118 & 15.4,14.5 &5.4,5.1& 14,16 & 15.4 & 32\\

CY~Tau & 17.2 &  14.1 &   5.7 & 71 & $\ldots$ & $\ldots$ & $\ldots$ &15.9 & 63\\

DF~Tau & 16.4 &  15.0 &   7.5 & 125 & 15.4&6.8& 99 & 16.3 & 27\\

DG~Tau &  15.8 &  14.4 &   17.8& 269& 14.5&6.4 & 86 & 16.3 & 20\\
          
DK~TauA & 15.7 &   15.6 &   7.8& 146 & 16.7&7.6 & 66 & 16.0 & 29\\

DL~Tau &  14.3 &   16.0 &   17.8& 275  & 16.1&6.9  & 158 & 15.8 & 19\\

DO~Tau & 17.5 &  18.0   &   6.8 & 106 & 17.2&6.7 & 44 &16.5 & 34\\

DR~Tau & 21.3 &  19.9   &   8.1 & 127 & 20.0&5.6 & 45 &  21.7& 25\\

GI~Tau &  17.6 &  18.0  &   8.0& 131 & 18.6&5.5 & 35 & 16.8 & 70 \\

GK~TauA &  18.2 &  18.6 &   7.2&  108 & 19.7&5.2 & 40 & 16.8 & 70\\

HN~TauA &  13.8\tablenotemark{c}  &  19.4 & 10.1 & 175 & 19.9&6.3& 75 & 22.8& 0.4\\

IP~Tau &  13.9\tablenotemark{d} &  15.2   & 6.2& 98 & 15.6&3.9 & 45& 15.8 & 23 \\

UX~TauA &  17.0 &  20.5 &   8.5 & 142 & 20.3&5.6 &39 & 19.0& 13\\

V819~Tau &  16.7 &  14.0&    5.7 & 72 & $\ldots$ & $\ldots$ & $\ldots$  & 16.0 & 37\\

\hline

TW~Hya & 10.1 & $\ldots$ &  $\ldots$ &  $\ldots$ &  $\ldots$ & $\ldots$  & $\ldots$ & & \\
\enddata
%% Text for table notes should follow after the \enddata but before
%% the \end{deluxetable}. Make sure there is at least one \tablenotemark
%% in the table for each \tablenotetext.
\tablecomments{Sources that have been observed more than once have multiple entries. Stellar RVs have a typical uncertainty of 1\,km/s (see Sect.~\ref{sect:analysis} for further details). Three dots are used to indicate a non detection. No entry in the CO columns means that there is no CO map at the location of the source.}
%\tablerefs{(1) This work; (2) Nguyen et al. 2012}
\tablenotetext{a}{The symbol $^\ast$ indicates that the stellar RV is computed from the Ca line (6439.07\,\AA) instead of being computed from the Li line.}
\tablenotetext{b}{This source was observed on two consecutive nights. Note the large change in the Li RV while the \sodium{} and \potassium{} centroids are essentially the same. Throughout the paper we will use 14.8\,km/s as the stellar RV because this value is in agreement with literature values (Hartmann et al. 1986 and  Nguyen et al. 2012).}
\tablenotetext{c}{Low signal to noise and shallow feature. The reported stellar RV is the mean of a thousand gaussian fits to the Li line as discussed in Sect.~\ref{sect:analysis}, its uncertainty is 3\,km/s.}
\tablenotetext{d}{This source was also observed in our 2012 campaign, the 2012 spectrum has a higher S/N than the 2006 spectrum. Note that both the stellar and the \sodium{} and \potassium{} RVs are shifted by $\sim$2\,km/s with respect to the 2012 values (see also Fig.~\ref{fig:NaKLi_multi}). This systematic shift suggests a difference in the overall wavelength calibration maybe due to centering the source in the slit. Because the 2012 stellar RV differs by only 0.4\,km/s from that reported in Nguyen et al.~(2012) we only use results from the 2012 spectrum in our analysis. 
}
\end{deluxetable}

\begin{deluxetable}{c c ccc ccc cc}
\tabletypesize{\scriptsize}
%\rotate
\tablecaption{2012 campaign: Heliocentric radial velocities, widths and EWs of selected lines.\label{tab:RV2012}}
\tablewidth{0pt}
\tablehead{
\colhead{} & \colhead{Stellar\tablenotemark{a}} & \multicolumn{3}{c}{Na}  & \multicolumn{3}{c}{K}  &\multicolumn{2}{c}{$^{12}$CO}\\
\colhead{Source} & \colhead{RV}  & \colhead{RV} & \colhead{FWHM} & \colhead{EW} & \colhead{RV} & \colhead{FWHM} & \colhead{EW} & \colhead{RV} & \colhead{S/N}\\
\colhead{ } & \colhead{[km/s]} & \colhead{[km/s]} & \colhead{[km/s]} & \colhead{[m\AA]} &\colhead{[km/s]} & \colhead{[km/s]} & \colhead{[m\AA]} &\colhead{[km/s]} & \colhead{}
}
\startdata
Anon~1 & 16.3  & 16.1 &   10.7  & 110 & $\ldots$ & $\ldots$ & $\ldots$ & 15.4 & 57\\

CI~Tau &  19.1\tablenotemark{b}  & 19.1 &   11.3 & 218 & 17.4,23.4 & 6.1,6.1 & 60 & 16.5 & 27\\

CoKu~Tau4 & 17.9,17.1  &  16.3,15.5  &  7.4,7.5 & 128,121 & 16.3,15.3 & 5.7,5.0 & 52,49 & 15.1 & 17\\

DH~TauA &  16.4  & 17.8 &    7.6 & 114 & 17.4 & 5.6 & 45 & 16.0 & 37 \\

DM~Tau &  18.5\tablenotemark{*}  & 21.5 &   6.8 & 90 & $\ldots$ & $\ldots$ & $\ldots$ & 21.8 & 1.4\\

DN~Tau &  17.6\tablenotemark{*}  & 18.1 &   9.1 & 142 & $\ldots$ & $\ldots$ & $\ldots$ & 16.6 & 44 \\

DS~TauA &  16.2  & 15.6 &   5.8 & 72 & $\ldots$ & $\ldots$ & $\ldots$ & 16.0 & 11\\
          
FM~Tau &  16.0 &  15.6 &    7.1 & 83 & 15.0& 3.2& 4 & 15.4 & 32\\

FZ~Tau & 18.0,18.1  & 19.8,19.9  &  8.4,8.7 & 140,138 & 20.4,20.6 & 5.9,6.7 & 81,89 & 16.4 & 83\\

GH~Tau &  18.4 & 19.6 &  7.5 & 111 & 20.2 & 5.1 & 42 & 16.5 & 35 \\

GM~Aur &  15.6  & 16.2 &  7.0 & 87 & $\ldots$ & $\ldots$ & $\ldots$ & 14.2 & 2.2 \\

GO~Tau & 17.2,17.5 &  18.1,18.3  &  6.7,6.0 & 106,86 & $\ldots$ & $\ldots$ &$\ldots$ & 16.6 & 28 \\

HBC~427 & 14.35\tablenotemark{c} &  16.2 & 7.1 & 123 & $\ldots$ & $\ldots$ & $\ldots$ & 14.7 & 4.9\\

HQ~Tau &  17.1 &  19.1 &   9.9 & 169 & 18.7,27.3 & 7.9,9.2 & 197 & 17.2 & 15 \\

IP~Tau &  16.1 &  17.3 &  6.8 & 86 & 17.7& 5.6& 30 & 15.8 & 23 \\

IT~TauA &  16.1 & 17.5 &  8.5 & 146 & 17.2 & 8.5& 95 & 15.4 & 30 \\

%% See comments in previous table as to why the source is removed!
%RXJ0405 & 7.1 & 1 \\ 
%%

UX~TauA &  16.5 &  20.7  &    8.3 & 150 & 19.9&5.4 & 36 &  19.0 & 13 \\

VY~Tau  &  17.1 &  20.9  &  7.8& 119 & $\ldots$ & $\ldots$   & $\ldots$ &  16.0 & 1.7 \\

V410~Tau & 16.0\tablenotemark{d}&  14.6   &   8.1  & 117 & 14.0&5.0& 4 & 16.1 & 74 \\

V710~TauA & 20.0 & 21.4 &  8.3 & 126 & 23.0 & 5.4& 32 & 18.8 & 28 \\

V773~Tau &  16.4\tablenotemark{c}  & 15.6 &   9.4 & 146 & 16.0&7.4& 34 & 15.3 & 34 \\

V836~Tau &  19.2\tablenotemark{*}  & 20.1 &   5.6& 57 & 18.8&5.7 & 31 & 18.1 & 13 \\

V1321~Tau & 21.3  & 22.1 &   7.5& 133& 22.7&3.1 & 12 & 21.5 & 1.9\\
   
V1348~Tau & 15.4\tablenotemark{*} & 22.9(?) &   8.4(?) & $\ldots$ & $\ldots$ & $\ldots$ & $\ldots$ & 24.0 & 1.6\\
\enddata
%% Text for table notes should follow after the \enddata but before
%% the \end{deluxetable}. Make sure there is at least one \tablenotemark
%% in the table for each \tablenotetext.
\tablecomments{Sources that have been observed more than once have multiple entries for the RV and FWHM. The reported optical RVs have a typical uncertainty of 1\,km/s while CO RVs of 0.2\,km/s (see Sect.~\ref{sect:analysis} for further details). Two components are present in the \potassium{} absorption line toward CI~Tau and HQ~Tau (see Fig.~\ref{fig:hqtau-citau}), hence we fit these profiles with the sum of two gaussian curves.}
%\tablerefs{(1) This work; (2) Nguyen et al. 2012; (3) Gontcharov 2006}
\tablenotetext{a}{The symbol $^\ast$ indicates that the stellar RV is computed from the Ca line (6439.07\,\AA) instead of being computed from the Li line.}
\tablenotetext{b}{Suspected spectroscopic binary (Nguyen et al. 2012).}
\tablenotetext{c}{Known spectroscopic binaries. We report and use the velocity of the barycenter of the systems as in Steffen et al.~(2011) for HBC~427 and Rivera et al. (2015) for V773~Tau.}
\tablenotetext{d}{Two components are visible in the Li line. The single gaussian fit we perform here is sensitive to how much of the continuum is included in the fit. The reported stellar RV is the mean of a thousand gaussian fits to the Li line as discussed in Sect.~\ref{sect:analysis}, its uncertainty is 3\,km/s.}
%\tablenotetext{\diamond}{Two components are present in the \potassium{} absorption line toward CI~Tau and HQ~Tau
%(see Fig.~\ref{fig:hqtau-citau}). The centroid and FWHM reported in the Table are for one gaussian fit. For %CI~Tau the two components are at velocities 18.2 and 23.1\,kms/s while for HQ~Tau they are at 20.1 and 27.0\,km/%s.}
\end{deluxetable}

%% If you use the table environment, please indicate horizontal rules using
%% Do not put multiple tabular environments within a single table.
%% The optional \label should appear inside the \caption command.

%% If the table is more than one page long, the width of the table can vary
%% from page to page when the default \tablewidth is used, as below.  The
%% individual table widths for each page will be written to the log file; a
%% maximum tablewidth for the table can be computed from these values.
%% The \tablewidth argument can then be reset and the file reprocessed, so
%% that the table is of uniform width throughout. Try getting the widths
%% from the log file and changing the \tablewidth parameter to see how
%% adjusting this value affects table formatting.

%% The \dataset{} macro has also been applied to a few of the objects to
%% show how many observations can be tagged in a table.

%% Tables may also be prepared as separate files. See the accompanying
%% sample file table.tex for an example of an external table file.
%% To include an external file in your main document, use the \input
%% command. Uncomment the line below to include table.tex in this
%% sample file. (Note that you will need to comment out the \documentclass,
%% \begin{document}, and \end{document} commands from table.tex if you want
%% to include it in this document.)


%% The following command ends your manuscript. LaTeX will ignore any text

\end{document}